\documentclass[pdflatex,sn-basic]{sn-jnl}

\usepackage{xcolor}
\jyear{2021}%

\theoremstyle{thmstyleone}%
\theoremstyle{thmstyletwo}%

\theoremstyle{thmstylethree}%

\def\prl{Physical Review Letters}
\def\apjl{Astrophysical Journal Letters}
\def\apj{Astrophysical Journal}
\def\apjs{Astrophysical Journal Supplementary}

\def\ssr{Space Science Reviews}
\def\jgr{Journal of Geophysical Research}
\def\grl{Geophysical Research Letters}
\def\planss{Planetary and Space Science}
\def\mnras{Monthly Notices of the Royal Astronomical Society}
\def\prd{Physical Review D}
\def\araa{Annual Review of Astronomy and Astrophysics}

\def\aap{Astronomy and Astrophysics}
\def\aapr{Astronomy and Astrophysics Reviews}
\def\nat{Nature}

\begin{document}

\title[Article Title]{Magnetic Reconnection and Associated Particle Acceleration in High-energy Astrophysics}


\author*[1]{\fnm{Fan} \sur{Guo}}\email{guofan@lanl.gov} 

\author[2]{\fnm{Yi-Hsin} \sur{Liu}}\email{yi-hsin.liu@dartmouth.edu}
\author*[3]{\fnm{Seiji} \sur{Zenitani}}\email{seiji.zenitani@oeaw.ac.at}
\author[4]{\fnm{Masahiro} \sur{Hoshino}}\email{hoshino@eps.s.u-tokyo.ac.jp}


\affil[1]{Los Alamos National Laboratory, Los Alamos, New Mexico 87545, USA}

\affil[2]{Department of Physics and Astronomy, Dartmouth College, Hanover, New Hampshire 03755, USA}

\affil[3]{Space Research Institute, Austrian Academy of Sciences, Schmiedlstra{\ss}e 6, 8042 Graz, Austria}

\affil[4]{Department of Earth and Planetary Science, The University of
Tokyo, Tokyo, 113-0033, Japan}


\abstract{Magnetic reconnection occurs ubiquitously in the universe and is often invoked to explain fast energy release and particle acceleration in high-energy astrophysics. The study of relativistic magnetic reconnection in the magnetically dominated regime has surged over the past two decades, revealing the physics of fast magnetic reconnection and nonthermal particle acceleration. Here we review these recent progresses, including the magnetohydrodynamic and collisionless reconnection dynamics as well as particle energization. The insights in astrophysical reconnection strongly connect to the development of magnetic reconnection in other areas, and further communication is greatly desired. We also provide a summary and discussion of key physics processes and frontier problems, toward a better understanding of the roles of magnetic reconnection in high-energy astrophysics.
}




\maketitle
\tableofcontents

\section{Introduction}\label{sec1}

Magnetic reconnection is a ubiquitous process that occurs in many space, solar, astrophysical, and laboratory systems. It was initially proposed to explain the fast energy release and acceleration of particles in space and astrophysical systems \citep[e.g.,][]{Parker1957,Sweet1958,Petschek1964}. During reconnection, magnetic topology changes lead to the rapid release of magnetic energy in highly conducting plasmas that cannot be explained by magnetic diffusion. Magnetic reconnection is now widely considered as a pivotal process for explosive energy release, high-energy particle acceleration and radiation in the universe \citep{Uzdensky2011SSRv,Hoshino2012,Arons2012,Blandford2017,Guo2020,Ji2022}. 

In high-energy astrophysics, magnetic reconnection can occur in pulsar wind nebulae (PWNe) and pulsar magnetosphere, relativistic jets of active galactic nuclei (AGNs) and gamma-ray bursts, accretion disks and coronae surrounding massive compact objects, as well as strong magnetic field regions in magnetars, etc. (see Section \ref{sec1.1} for a more extensive discussion). The reconnection region is expected to be much larger than the kinetic scale, so a magnetohydrodynamic (MHD) description is necessary. However, many regimes of high-Lundquist-number magnetic reconnection show multiple X-lines and flux ropes (islands in 2D) developing as the secondary tearing instability is active in a current layer. Kinetic
processes are important in a collisionless system \citep{Daughton2007,Guo2015,Sironi2016} or when a hierarchy of collisional plasmoids \citep{Biskamp1986,Shibata2001,Loureiro2007,Bhattacharjee2009,Uzdensky2010} develop kinetic-scale current layers that may trigger collisionless reconnection \citep{Daughton2009,Ji2011,Stanier2019}. 
 Magnetic reconnection has been proposed as a mechanism to explain a broad range of high-energy astrophysical phenomena and radiation signatures. This includes high-energy radiation flares and their fast variability and polarized emission signature \citep{Zhang2018,Zhang2020,ZhangH2021b,Zhang2022,ZhangH2021,Petropoulou2016}, as well as emissions from the accretion flows recently observed by the Event Horizon Telescope (EHT)  \citep{Ripperda2020} and fast radio bursts (FRBs) \citep{Philippov2019,mahlmann22a}.

In many high-energy astrophysical systems, it is often estimated that magnetic reconnection, if occurs, will proceed in a magnetically dominated environment. The parameter for measuring the dominance is the so-called magnetization parameter\footnote{In the nonrelatvistic case, it is more adequate to use plasma $\beta$ or $\sigma_T = B^2/(6 \pi nT)$ to measure the magnetic field dominance. See Drake et al. (2024, this issue)}: 
\begin{equation}
\sigma = B^2/(4 \pi w) 
\end{equation}

\noindent with the enthalpy density $w = nmc^2 + [\Gamma_a/(\Gamma_a-1)]P$. Here $B$ is the magnitude of the magnetic field, $n$ is the proper plasma density, $c$ is the speed of light, $\Gamma_a$ is the ratio of specific heats, and $m$ and $P$ are the rest mass and proper pressure of the plasma particles under study, which can be the electron-positron pairs or electron-proton pairs, or a mixture of different species. It is expected that in many situations, $\sigma$ can be much larger than unity and the Alfv\'en speed $v_A/c = \sqrt{\sigma/(\sigma+1)}$ is close to the speed of light.
In terms of the energy budget, $\sigma_r = B_r^2/(4 \pi w)$ is the ratio between the potential free energy carried by the Poynting flux to the particle energy density flux into the reconnection region ($B_r$ represents the reconnecting field component). Analytical theories and MHD simulations have been developed to understand relativistic magnetic reconnection (Section \ref{sec2}). Particle-in-cell (PIC) simulations have greatly enhanced our understanding of relativistic magnetic reconnection in the high-$\sigma$ regime. Magnetic reconnection has been found to support a fast reconnection rate $R\sim 0.1-0.3$, indicating fast energy release \citep{Lyubarsky2005,Liu2015,Liu2017,Liu2020,Werner2018,Goodbred2022}. It has been shown to efficiently convert a sizeable fraction of the magnetic energy, leading to a strong particle energization. These reconnection layers are shown to strongly accelerate particles to high energy, leading to power-law energy spectra $f = dN/dE = f_0 E^{-p}$ with spectral index approaching $p\sim1$ for large $\sigma$ \citep{Zenitani2001,Sironi2014,Guo2014,Guo2015,Werner2016}. The key physics and observational implications of these results are being actively studied. We discuss collisionless reconnection physics, and plasma heating and particle acceleration in relativistic magnetic reconnection in Sections \ref{sec3} and \ref{sec4}, respectively. Note that this magnetically-dominated regime, collisionless shocks may be inefficient in dissipating magnetically dominated flows and accelerating energetic particles \citep[e.g.,][]{Guo2015,Sironi2015}, magnetic reconnection is the primary candidate for dissipating and converting magnetic energy into relativistic particles and subsequent radiation. 

While the initial studies focused on electron-positron plasmas, recent studies have extended into electron-proton plasmas. For high-$\sigma$ regime ($\sigma \sim \sigma_i \gg1$), the behavior of reconnection is similar to the pair plasma case, and both electrons and protons are efficiently accelerated \citep{Guo2016,Zhang2018}. Recent studies have also studied the so-called trans-relativistic regime, where $\sigma_i < 1$ but  $\sigma_e \gg 1$, which can lead to strong electron energization \citep{Werner2018,Ball2018,Kilian2020}. The transrelativistic regime is also a bridge for connecting the highly relativistic regime ($p \gtrsim 1.5$) with the nonrelativistic reconnection studies ($p \sim 4$) \citep{Dahlin2014,Li2021,Li2019,Zhang2021} (See \citet{Oka2023} and Drake et al. in this collection).
Traditionally, relativistic magnetic reconnection in the collisionless regime (usually pair plasma) has been less of a focus compared to the nonrelativistic cases. The connection between astrophysical reconnection and reconnection in other fields of research is strong, and communications should be continuously encouraged. 

In this paper, we review the recent progress in understanding magnetic reconnection in the relativistic magnetically dominated regime. We discuss the astrophysical systems that host magnetic reconnection and how magnetic reconnection may explain high-energy astrophysics observations in Section \ref{sec1.1}. We introduce a list of outstanding problems in relativistic magnetic reconnection in Section \ref{sec1.2}. Section \ref{sec2} discusses MHD models of relativistic magnetic reconnection. In Section \ref{sec3}, we discuss the reconnection structure and rate, as well as the generalized Ohm's law in a collisionless plasma. Section \ref{sec4} discusses the heating and acceleration due to relativistic magnetic reconnection. We will discuss basic acceleration mechanisms, the main features of nonthermal particle energy spectra, and the physics that determines the spectra, such as the low-energy injection and energy partition, power-law formation, and high-energy rollover. Section \ref{sec5} provides a final remark and possible future directions.

\subsection{Where and how magnetic reconnection may happen in astrophysical systems?} \label{sec1.1}

Relativistic outflows such as pulsar winds and relativistic jets are launched with energy carried largely in the form of Poynting flux \citep{Coroniti1990,Li2006,Spruit2010}. However, the magnetic energy in the flows must be eventually converted into energies in thermal and nonthermal particles to power the observed emission signatures. There is ample observational evidence indicating that astrophysical systems with strong magnetic field dissipation support efficient particle acceleration and high-energy radiation \citep{Abdo2011,Tavani2011,Abeysekara2017,Zhang2015,Ackermann2016}. Magnetic reconnection is a leading mechanism that explains this underlying process. 

\begin{figure}[ht]%
\centering
\includegraphics[width=\textwidth]{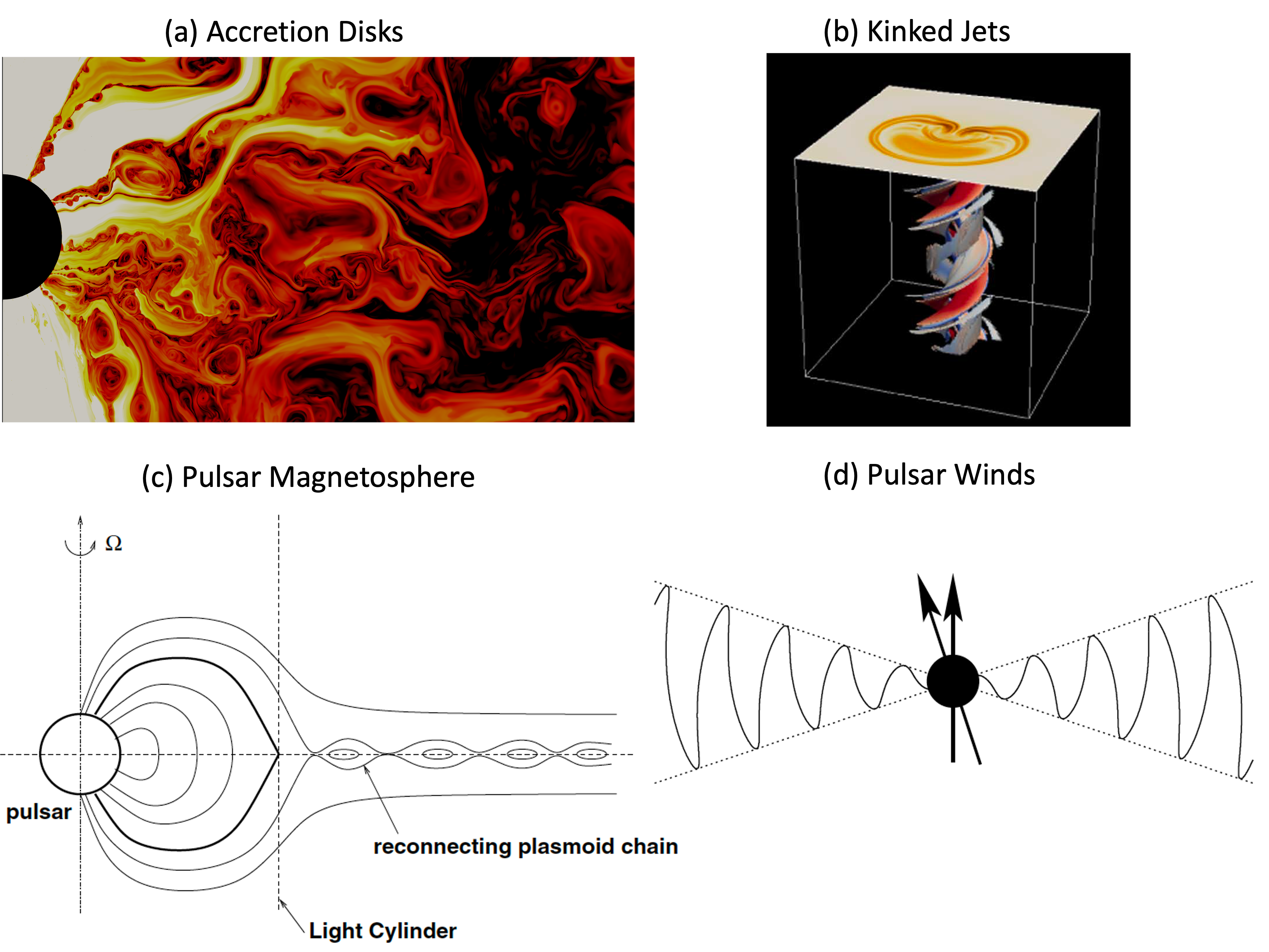}
\caption{Several examples of high-energy astrophysical systems that host magnetic reconnection: a) Accretion disks of black holes  (Figure created by Bart Ripperda using global general relativistic MHD simulations \citep{Ripperda2020}), b) Kinked relativistic jets \citep{Zhang2018}, c) pulsar magnetosphere \citep{Uzdensky2014}, and d) pulsar wind nebulae \citep{Kirk2003}. b), c), and d) are reproduced by permission of the AAS.\label{figure:fig1}}
\end{figure}

Fig. \ref{figure:fig1} summarizes several systems and environments where magnetic reconnection can be found. In pulsar wind nebulae (PWNe), the antiparallel component of the pulsar dipole field gives rise to a current sheet. The fast rotation of the obliquely oriented dipole leads to the so-called striped wind, where magnetic field directions reverse alternatively and can support magnetic reconnection in the equatorial region. Magnetic reconnection may occur starting from the pulsar magnetosphere \citep{Arons2012,Uzdensky2014,Philippov2022} to the pulsar wind \citep{Lyubarsky2001,Kirk2003}, and in the downstream of the termination shock driven by the shock compression \citep{Sironi2011,Lu2021}. A similar situation can happen in relativistic jets, so-called striped jets \citep{ZhangH2021}.  For the pulsar wind and pulsar magnetosphere, it is known that close to the neutron star, $\sigma$ is very high. In fact, how the magnetic field energy is dissipated before pulsar winds reach the termination shock is one of the unsolved science questions in the physics of PWNe (``$\sigma$ problem''). Magnetic reconnection in the wind and/or at the termination shock is the key process for solving the problem \citep{Coroniti1990,Kirk2003,Sironi2011,Lu2021,Zrake2016}. In the more extreme limit, the magnetic field in magnetars can be strong enough to trigger QED effects \citep{Uzdensky2011SSRv}. Current sheets may develop during neutron star coalescence as their magnetospheres interact, leading to magnetic reconnection \citep{Palenzuela2013}. Magnetic reconnection in highly magnetized regimes of pulsar magnetosphere has been proposed to drive fast radio bursts \citep[e.g.,][]{Philippov2019,mahlmann22a}.

Close to black holes, magnetic reconnection has also been proposed to explain emissions from accretion disks and relativistic jets. Magnetic reconnection can happen in the accretion disks and their coronae \citep{Hoshino2015,Ball2018,Ripperda2020,Ripperda2022,Nathanail2022,Lin2023,Hakobyan2023}. 
As kink instability in jets evolves nonlinearly \citep{Zhang2017,Bodo2021}, it can develop field-line reversals and support magnetic reconnection\footnote{We note that the kink instability itself can also support magnetic energy conversion and particle acceleration \citep{Alves2018}.}. 
In gamma-ray burst models, collisions of relativistic outflows with different magnetic field orientations are discussed \citep{Zhang2011}. Magnetic reconnection may provide the efficient energy dissipation needed in explaining gamma-ray bursts \citep{Zhang2011,McKinney2012}. Magnetic reconnection may explain polarized radiation signatures \citep{Zhang2018}. Recent general relativistic (GR) MHD simulations show that magnetic reconnection can happen in the accretion disk. Especially, there exists a low-$\beta$ corona that may have the right condition for strong particle acceleration and powering the nonthermal emission \citep{Ripperda2020,Yang2022}.

A number of radiation scenarios have been developed to explain high-energy emissions from astrophysical objects.
One of the strong motivations to consider magnetic reconnection is the Crab flare, where particles are likely to be accelerated to $10^{15}$ eV, and it is difficult for a shock model to explain it and avoid synchrotron cooling. Uzdensky and colleagues have proposed that extreme acceleration can happen in the reconnection region with a weak magnetic field ($E>B$) and exceed the ``burnoff limit'' \citep{Uzdensky2011,Cerutti2013}. Whether this can be realized is an active field of research. In particular, whether the global observable effect or only beaming at kinetic scale can be observed is under investigation \citep{Mehlhaff2020}.
Reconnection has also been used to explain the high energy emission and fast radiation variability in blazars from black holes as the reconnection outflows can Lorentz boost the radiation \citep{Giannios2009,Agarwal2023}. However, kinetic simulations observe only small regions/structures close to the upper limit of the Alfv\'en Lorentz factor, and therefore it is still uncertain if the ``minijet model'' can actually work \citep{Guo2015,Sironi2016}. In addition, the condition for obtaining these minijets requires a very weak guide field \citep{Liu2017}. How to achieve these conditions is still unclear and requires further study. Magnetic reconnection may also provide an explanation for time-dependent emissions \citep{Petropoulou2016} and polarized emissions in blazars \citep{Zhang2018,Zhang2021}. However, it is unclear if these radiation features are still preserved in the 3D reconnection, where the flux ropes are shown to be highly dynamic and can easily be disrupted \citep{Guo2021,Guo2016b}.

It is also interesting to mention that magnetic reconnection can happen in the foreshock region for high Mach number shocks \citep{Matsumoto2015}. Several numerical simulations have indicated that a number of current sheets can be generated by the ion Weibel instability during the interaction of incoming and reflected ions at the foreshock region \citep{Kato2008,Spitkovsky2008,Burgess2016}, and it is suggested that magnetic reconnection may play an important role on electron heating and acceleration as the so-called shock injection process into the first-order Fermi acceleration \citep{Bohdan2020,Amano2022}.

\subsection{Key Physics Issues} \label{sec1.2}

Before getting into the detailed discussion, we close this section by discussing a list of frontier physical problems that are currently undergoing active studies in relativistic magnetic reconnection. 

The {\it Rate Problem and Reconnection Dynamics} is one of the long-standing problems in magnetic reconnection and is concerned with how fast magnetic reconnection proceeds. Previous analytical studies have made various predictions on the rate and reconnection dynamics in the relativistic regime. While the Sweet-Parker-like model cannot support fast reconnection, recent MHD simulations have shown relativistic Petschek reconnection and plasmoid-dominated reconnection with the rate insensitive to the Lundquist number.
Kinetic simulations and analysis have shown the rate can be up to $R\sim 0.3$. Figuring out how fast magnetic energy is dissipated can explain astrophysical magnetic energy release. Learning the reconnecting electric field helps us to understand the upper limit of particle acceleration that magnetic reconnection can explain. In Sections \ref{sec2} and \ref{sec3}, we will discuss recent studies of relativistic reconnection dynamics via fluid approach and fully kinetic approach, respectively.

The {\it Particle Heating and Acceleration Problem} is to understand how magnetic energy is converted and partitioned into thermal and nonthermal particles, and how a population of particles are accelerated to high energy. An important goal of the particle energization problem is to build a complete understanding of the physics processes involved and predict the resulting distributions of energetic particles during relativistic magnetic reconnection. A major achievement over the past decade is the robust evidence that magnetic reconnection is a source of nonthermal particles, producing clear power-law particle energy distribution. Competing theories have been proposed based on various reconnection models, but there is still no general consensus on the origin of nonthermal distributions observed during magnetic reconnection. In Section \ref{sec4}, we summarize recent progress on particle acceleration in relativistic reconnection.

{\it 3D Reconnection and Effects of Turbulence}: The main issue here is to understand how reconnection properties and associated particle acceleration change compared to 2D reconnection. A closely related problem is the role of turbulence, either externally driven or self-generated, during the reconnection process. Recently, 3D kinetic simulations and high-Lundquist-number MHD simulations (with and without preexisting turbulence) have been carried out. However, it is still unclear if and how reconnection physics is strongly influenced by turbulence. Meanwhile, there seems to be promising evidence suggesting that particle acceleration in 3D becomes substantially different. We will discuss the 3D effects and role of turbulence in the following sections as we discuss individual topics.

{\it The Onset Problem:} Most studies have been focusing on pre-existing current sheets. However, the current sheet needs to form in the first place, which is a dynamic and sometimes prolonged process \citep[e.g.,][]{Uzdensky2016b,Tenerani2016,Huang2017,Comisso2017,Lyutikov2017}. In addition, how magnetic energy is accumulated and stored prior to the onset is unknown. Thus, it is important to include the formation of current sheets and their effect. In addition, learning how reconnection onsets helps us understand the conditions that lead to explosive energy release in astrophysics.

{\it Multiscale Problem and Reconnection in Global Models:} Reconnection in most astrophysical problems must involves both MHD scales and kinetic scales. While MHD models offer a basic description of large-scale magnetic reconnection, studies have shown kinetic effects are essential. Developing a self-consistent treatment that includes multi-scale effects is of central importance. In addition, developing global particle acceleration models is essential for describing particle acceleration in realistic systems and connecting with observations.

The {\it Radiation Problem} is to study the effect of radiation cooling (on reconnection structure), pair production, radiation pressure, etc. In addition, there is strong interest in modeling the radiation signature to explain observations. This is not the focus of this review, but we refer interested readers to the recent papers by \citet{Jaroschek2009} and \citet{Uzdensky2011SSRv}.

\section{Fluid Simulations of Relativistic Reconnection}\label{sec2}




\subsection{Early theories}

\citet{Blackman1994} presented
the first theoretical models of steady relativistic reconnection.
By constructing relativistic Sweet--Parker and Petschek models,
they pointed out that plasma density in the outflow region increases due to Lorentz contraction.
Since this, in turn, requires relativistically fast plasma inflow, 
they claimed that the reconnection rate can approach unity, $\sim \mathcal{O}(1)$.
\citet{Lyutikov2003} further examined the relativistic Sweet--Parker model and
suggested that reconnection outflow speed may exceed inflow Alfven speed.
Later, \citet{Lyubarsky2005} developed theoretical models of
relativistic Sweet--Parker and Petschek reconnection,
taking compressibility into account.
In the relativistic regime, he pointed out that
the internal energy density in the outflow region is high enough
to increase the effective plasma inertia.
For this reason, he claimed that the reconnection outflow is only mildly relativistic.
In addition, considering the full momentum balance in Petschek's model,
he predicted that the opening angle of slow shocks will be smaller.
This leads Lyubarsky to argue that the reconnection rate remains on the order of $\mathcal{O}(0.1)$,
because it is difficult to transport energy through a narrow outflow exhaust.

\subsection{Basic equations}
\label{RRMHD}

To validate the theories, several numerical models have been developed over the past decades.
Roughly speaking, two kinds of numerical models have been used, 
relativistic resistive magnetohydrodynamics (RRMHD) and
relativistic multifluid dynamics.

The RRMHD model was developed by \citet{naoyuki06} and by \citet{Komissarov07}.
Combining relativistic fluid equations, Ohm's law, and Maxwell's equations,
they have organized the following set of RRMHD equations,
\begin{align}
\partial_t (\Gamma \rho) + \nabla \cdot (\rho {\bf U}) &= 0, \\
\partial_t ( \Gamma w{\bf U} + {\bf E}\times{\bf B} )
+ \nabla \cdot \Big( ( P + \frac{B^2+E^2}{2} ) \mathbb{I}
+ w {\bf U}{\bf U}
- {\bf B}{\bf B} - {\bf E}{\bf E} \Big)
&= 0, \\
\partial_t ( \Gamma^2w-P + \frac{ {B}^2+{E}^2 }{2} )
+ \nabla \cdot ( \Gamma w{\bf U} + {\bf E}\times{\bf B} ) &= 0, \\
\partial_t{\bf B} + \nabla \times {\bf E} = 0, ~~~~~~
\partial_t{\bf E} - \nabla \times {\bf B} &= -{\bf J},
\label{eq:Max} \\
\partial_t{\bar{\rho}_c} + \nabla \cdot {\bf J} &= 0,
\\
\label{eq:ohm}
\Gamma \Big( {\bf E} + {\bf V}\times{\bf B} - ({\bf E}\cdot{\bf V}) {\bf V} \Big)
&= \eta ( {\bf J} - \bar{\rho}_c {\bf V} )
\end{align}
Here, we have used Lorentz--Heaviside units with $c=1$.
In equations, $\Gamma \equiv 1/\sqrt{1-(V/c)^2}$ is the Lorentz factor,
$\rho$ is the proper density,
${\bf U}=\Gamma{\bf V}$ is the 4-vector,
$w$ is the relativistic enthalpy,
$P$ is the proper pressure, $\mathbb{I}$ is the identity matrix,
and $\bar{\rho}$ is the observer-frame charge density.
The internal energy in the relativistic enthalpy is often approximated by
a simple equation of state with the adiabatic index $\Gamma_a=4/3$.
In such a case, the enthalpy is given by $w=\rho c^2+[\Gamma_a/(\Gamma_a-1)]P$. 

This form of the Ohm's law may not be familiar to all the readers (Eq.~\eqref{eq:ohm}; \citet{Komissarov07}).
In relativity, we consider the covariant electric field $e^{\mu} \equiv [ \Gamma ({\bf E}\cdot {\bf V}), \Gamma ( {\bf E} + {\bf V}\times{\bf B}) ]$ and the electric current $J^{\mu} \equiv (\bar{\rho}_c, {\bf J})$ in the four-vector form. The latter is further split into the conduction current $j^{\mu}$ and the convection current $\rho_c {U}^{\mu}$, a projection of the motion of the rest-frame charge $\rho_c$.
The Ohm's law relates the electric field and the conduction current,
i.e., $e^{\mu} = \eta j^{\mu} = \eta (J^{\mu}-\rho_c U^{\mu})$.
In the three-vector form, this gives
\begin{align}
\Gamma \Big( {\bf E}\cdot{\bf V} \Big)
&= \eta ( {\bf \bar{\rho}_c} - \rho_c {\Gamma} )
\label{eq:rohm1}
\\
\Gamma \Big( {\bf E} + {\bf V}\times{\bf B} \Big)
&= \eta ( {\bf J} - \rho_c \Gamma{\bf V} )
\label{eq:rohm2}
\end{align}
The Ohm's law (Eq.~\eqref{eq:ohm}) is obtained
by subtracting Eq.~\eqref{eq:rohm1} $\times {\bf V}$ from Eq.~\eqref{eq:rohm2}.
The ${\bf E}\cdot{\bf V}$ term is purposely added to make the Ohm's law simulation-friendly.
\citet{naoyuki06} have used a different but equivalent form.

Practically, Eqs.~\eqref{eq:Max} are often replaced by the following equations \citep{Komissarov07}.
\begin{align}
\partial_t{\bf B} + \nabla \times {\bf E} + \nabla \Phi = 0, ~~~~~~
\partial_t{\bf E} - \nabla \times {\bf B} + \nabla \Psi = -{\bf J},
\label{eq:div_cleaning1}
\\
\partial_t \Phi + \nabla \cdot {\bf B} = -\kappa \Phi, ~~~~~~
\partial_t \Psi + \nabla \cdot {\bf E} = \rho_c -\kappa \Psi,
\label{eq:div_cleaning2}
\end{align}
Here, $\Phi, \Psi, \kappa$ are virtual potentials and the decay coefficient
in order to reduce numerical errors in $\nabla\cdot{\bf B}$ and $\nabla\cdot{\bf E}$
by the so-called hyperbolic divergence cleaning method \citep{Munz2000,Dedner2002}.
In the RRMHD equations, the electric field is known to be very stiff,
when the resistivity $\eta$ is low.
To deal with these issues, various schemes have been developed
such as the operator splitting \citep{Komissarov07},
implicit schemes \citep{Palenzuela2009,Dumbser2009,Mignone2019}, and
a method of characteristics \citep{Takamoto2011}.
Other extensions include the Galerkin method \citep{Dumbser2009} and generalized equation of states (EoSs) \citep{Mizuno2013}.

Another approach is to use relativistic multifluid equations.
\citet{Zenitani2009a,Zenitani2009b} have proposed the following relativistic multifluid equation systems.
For electron-positron two-fluid plasma,
this equation system is also known as relativistic two-fluid electrodynamics.
They consist of relativistic fluid equations and Maxwell equations.
\begin{eqnarray}
\label{eq:cont}
\partial_t ( \Gamma_p n_p ) &=& -\nabla \cdot (n_p {\bf U}_p), \\
\label{eq:mom}
\partial_t \Big( { \Gamma_p w_p {\bf U}_p } \Big)
&=& -\nabla \cdot \Big( { w_p {\bf U}_p{\bf U}_p } + P_p \mathbb{I} \Big)
+ \Gamma_p n_p q_p ({\bf E}+{\bf V}_p\times{\bf B})
- \tau n_p n_e ({\bf U}_p-{\bf U}_e), \\
\label{eq:ene}
\partial_{t} \Big(\Gamma_p^2 w_p - P_p \Big)
&=& -\nabla \cdot ( \Gamma_p w_p {\bf U}_p ) + \Gamma_p n_p q_p ({\bf V}_p\cdot{\bf E})
- \tau n_p n_e ({\Gamma}_p-{\Gamma}_e),\\
\label{eq:cont2}
\partial_t ( \Gamma_e n_e ) &=& -\nabla \cdot (n_e {\bf U}_e), \\
\label{eq:mom2}
\partial_t \Big( { \Gamma_e w_e {\bf U}_e } \Big)
&=& -\nabla \cdot \Big( { w_e {\bf U}_e{\bf U}_e } + P_e \mathbb{I} \Big)
+ \Gamma_e n_e q_e ({\bf E}+{\bf V}_e\times{\bf B})
- \tau n_p n_e ({\bf U}_e-{\bf U}_p), \\
\label{eq:ene2}
\partial_{t} \Big(\Gamma_e^2 w_e - P_e \Big)
&=& -\nabla \cdot ( \Gamma_e w_e {\bf U}_e ) + \Gamma_e n_e q_e ({\bf V}_e\cdot{\bf E})
- \tau n_p n_e ({\Gamma}_e-{\Gamma}_p), \\
\partial_t {{\bf B}} &=& - \nabla \times {\bf E},
~~~~~~~~
\partial_t {{\bf E}} = \nabla \times {\bf B} - 4\pi ( q_p n_p {\bf U}_p+q_e n_e {\bf U}_e ).
\end{eqnarray}
In these equations,
the subscript $p$ indicates positron properties (and $e$ for electrons),
$n$ is the proper number density, and $\tau$ is a friction coefficient.
For simplicity, the rest mass $m$ and the light speed $c$ are set to $1$.
Note that interspecies friction terms and the $\tau$ parameter are added to
the right-hand sides of the momentum and energy equations \citep{Zenitani2009b}.
The friction terms control the collision between the two species.
Therefore, they (and fluid inertial terms) act as a resistivity.
By setting $\tau$ to small,
the friction terms are essentially unused in the inflow region,
but we sometimes increase $\tau$ near the X-line
when we desire a localized resistivity. 
Similarly, divergence cleaning potentials are often used to
improve the numerical accuracy (Eqs. \eqref{eq:div_cleaning1} and \eqref{eq:div_cleaning2}).
Numerical schemes to better solve these equations have been actively developed
over the years \citep{barkov14,Balsara2016,Amano2016}.

\subsection{Relativistic Petschek reconnection}
\label{RPR}

\begin{figure}
\centering
\includegraphics[width=0.45\columnwidth]{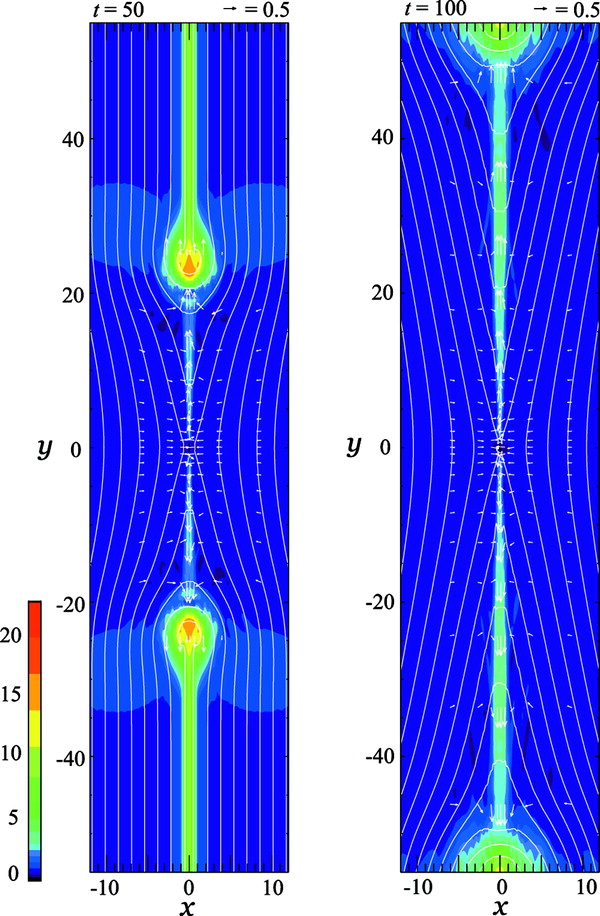}
\caption{
Evolution of plasma density from RRMHD simulation with $\sigma=4.0$ and localized resistivity in the reconnection plane.
The solid lines and the arrows show magnetic field lines and velocity vectors.
[Adapted from \citet{naoyuki06}, reproduced by permission of the AAS].
\label{fig:naoyuki06}
}
\end{figure}

\begin{figure}
\centering
\includegraphics[width=0.6\columnwidth]{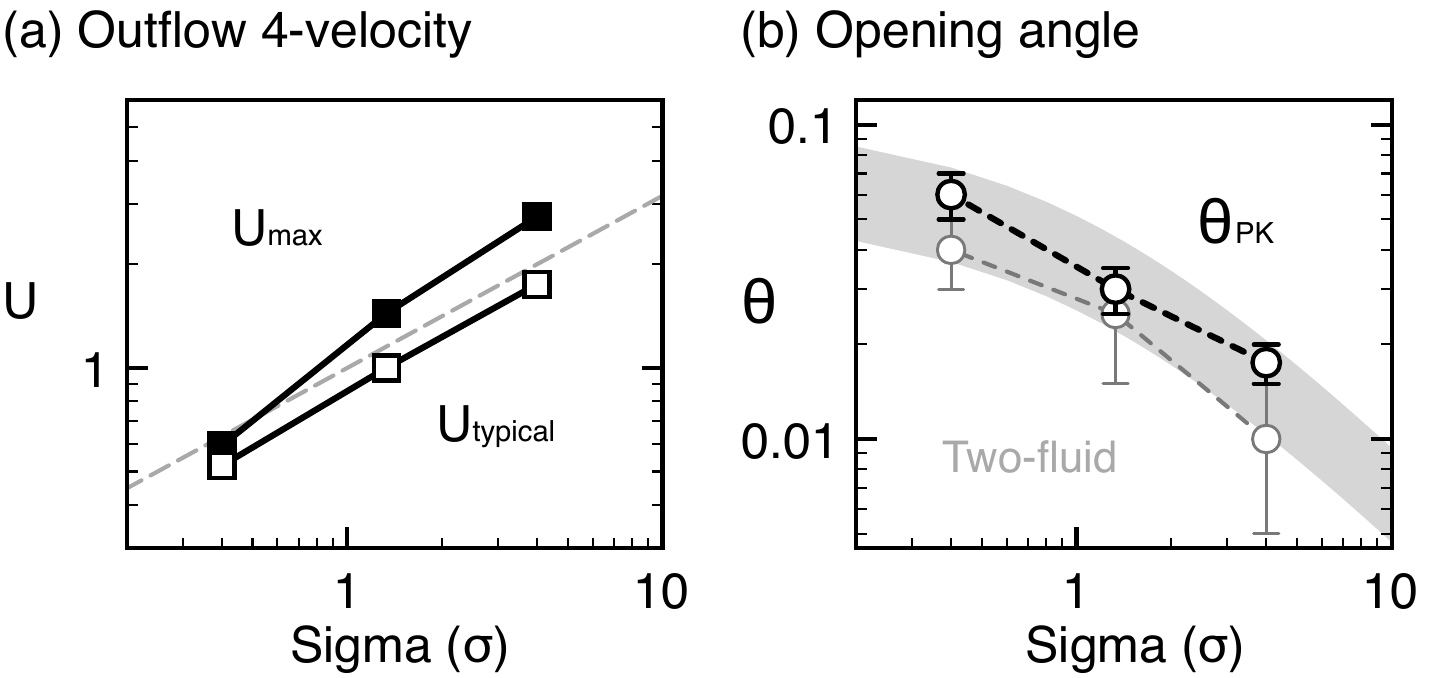}
\caption{
(a) Maximum and typical outflow 4-velocities $U$ as a function of the inflow $\sigma$.
The dashed line indicates an Alfv\'{e}n speed in Eq.~\eqref{eq:Alfven}.
(b) Opening angles of the Petschek slow shocks
in RRMHD (black line) and in two-fluid (gray) simulations.
The shadow indicates a predicted scaling of $\propto (1+\sigma)^{-1}$
[Adapted from \citet{Zenitani2010}, reproduced by permission of the AAS].
\label{fig:zeni10}
}
\end{figure}

\begin{figure}
\centering
\includegraphics[width=0.8\columnwidth]{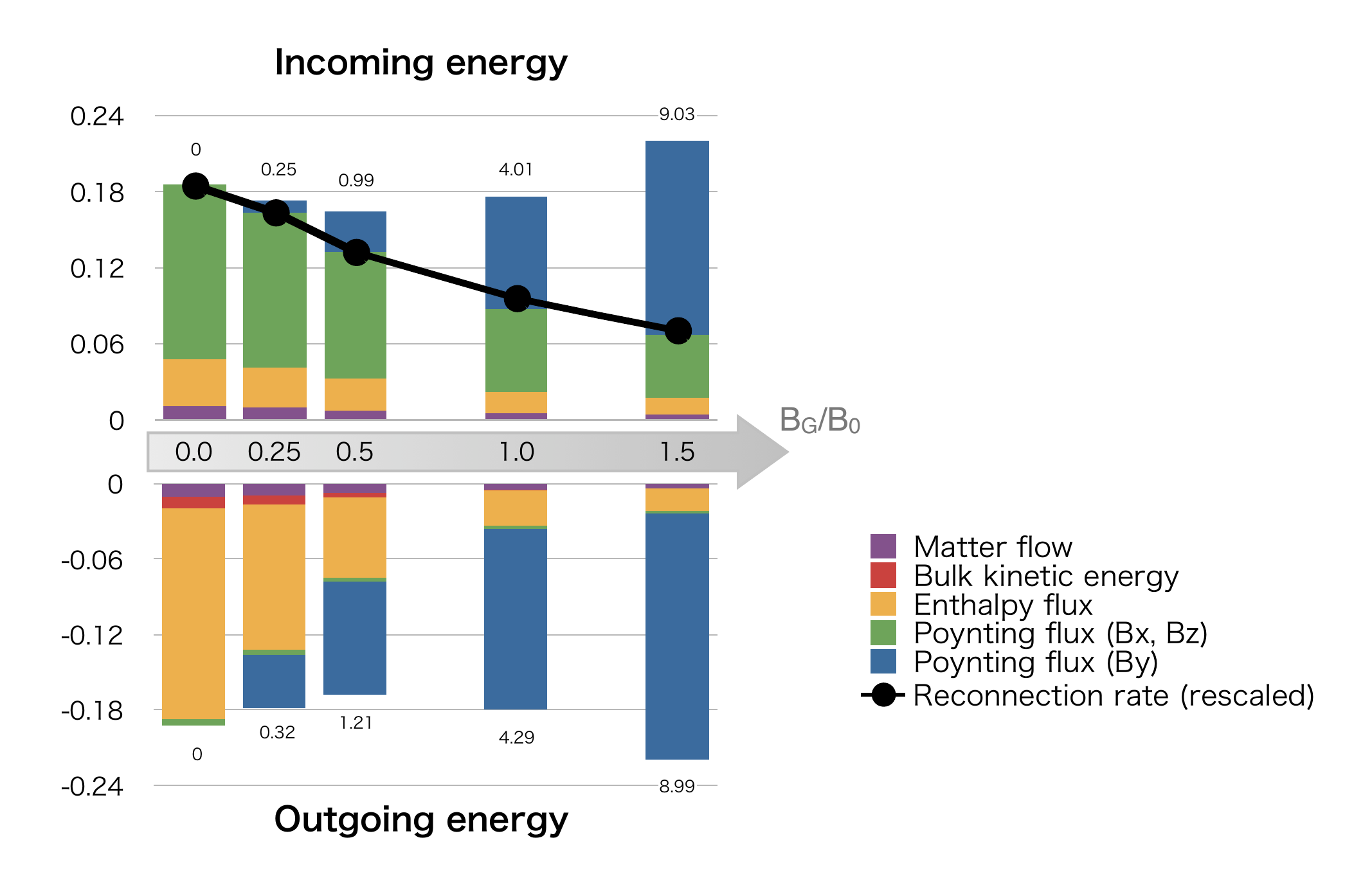}
\caption{
The incoming and outgoing energy fluxes around the reconnection region.
Note that we respect the coordinate system in the original articles.
In this case, the antiparallel magnetic field component is along the x direction,
the inflow is along z, and the guide field is along y.
The guide field Poynting flux ($B_y$), the rest part of the Poynting flux ($B_x$, $B_z$), the plasma enthalpy flux, the bulk kinetic energy, and the matter flow are presented.
The black curve indicates a rescaled reconnection rate.
[Adapted from \citet{Zenitani2009b}, reproduced by permission of the AAS]
\label{fig:zeni09b}}
\end{figure}

Using RRMHD equations, \citet{naoyuki06} have pioneered
the MHD-scale evolution of relativistic magnetic reconnection.
They have assumed a spatially localized resistivity $\eta=\eta(x,y)$
in the relativistic Ohm's law (Eq. \ref{eq:ohm}),
and then they have obtained a well-developed picture of
relativistic Petschek reconnection.
As shown in Fig. \ref{fig:naoyuki06},
a narrow reconnection jet extends from the central reconnection point.
The reconnection jet is surrounded by a pair of slow shocks,
similar to nonrelativistic Petschek reconnection \citep{Petschek1964}. 
It appears that
the outflow channel is much narrower than in the nonrelativistic case.
These features were further examined by subsequent studies
by two-fluid \citep{Zenitani2009a,Zenitani2009b} and the RRMHD simulations \citep{Zenitani2010,Zanotti2011}.

It has been found that
the typical outflow speed is approximated by
the upstream Alfv\'{e}n speed \citep{Zenitani2010},
\begin{align}
V_{\rm out} \approx c_{\rm A,up} = c~\sqrt{\frac{\sigma}{1+\sigma}},
\label{eq:Alfven}
\end{align}
where $\sigma \equiv B^2/w$ (in the Lorentz--Heaviside units) is the magnetization parameter in the upstream region.
This relation in the 4-velocity form is indicated by the white squares and the dashed line in Fig.~\ref{fig:zeni10}(a).
Also, the opening angle of the Petschek outflow becomes
narrower and narrower as the system becomes relativistic,
as confirmed in Fig.~\ref{fig:zeni10}(b).
These results are in excellent agreement with
theoretical predictions by \citet{Lyubarsky2005}.
Numerical simulations have revealed that
the reconnection rate is $\mathcal{R} \sim \mathcal{O}(0.1)$ or even faster.
Theoretically, such a fast reconnection was questioned before
because the outflow channel may be too narrow to
eject a sufficient amount of energy from the reconnection region \citep{Lyubarsky2005}.
Based on the numerical results in the relativistic two-fluid model \citep{Zenitani2009b},
we explain this logical gap in the following way.
Fig. \ref{fig:zeni09b} presents
the composition of the incoming and outgoing energy flow
during the quasi-steady stage of reconnection.
In the antiparallel (leftmost) case with $B_{\rm G}=0$,
it has been found that
the energy is mostly carried away
in the form of the relativistic enthalpy flux,
$\sim \sum_{i=p,e} 4\Gamma^2 P {\bf V}$,
which was often overlooked by the earlier theories.
In other words,
since the enthalpy flux can carry a huge amount of energy per unit rest mass,
it allows fast reconnection even through the narrow outflow channel.

\subsection{Relativistic Sweet--Parker reconnection}

\begin{figure}
\centering
\includegraphics[width=0.8\columnwidth]{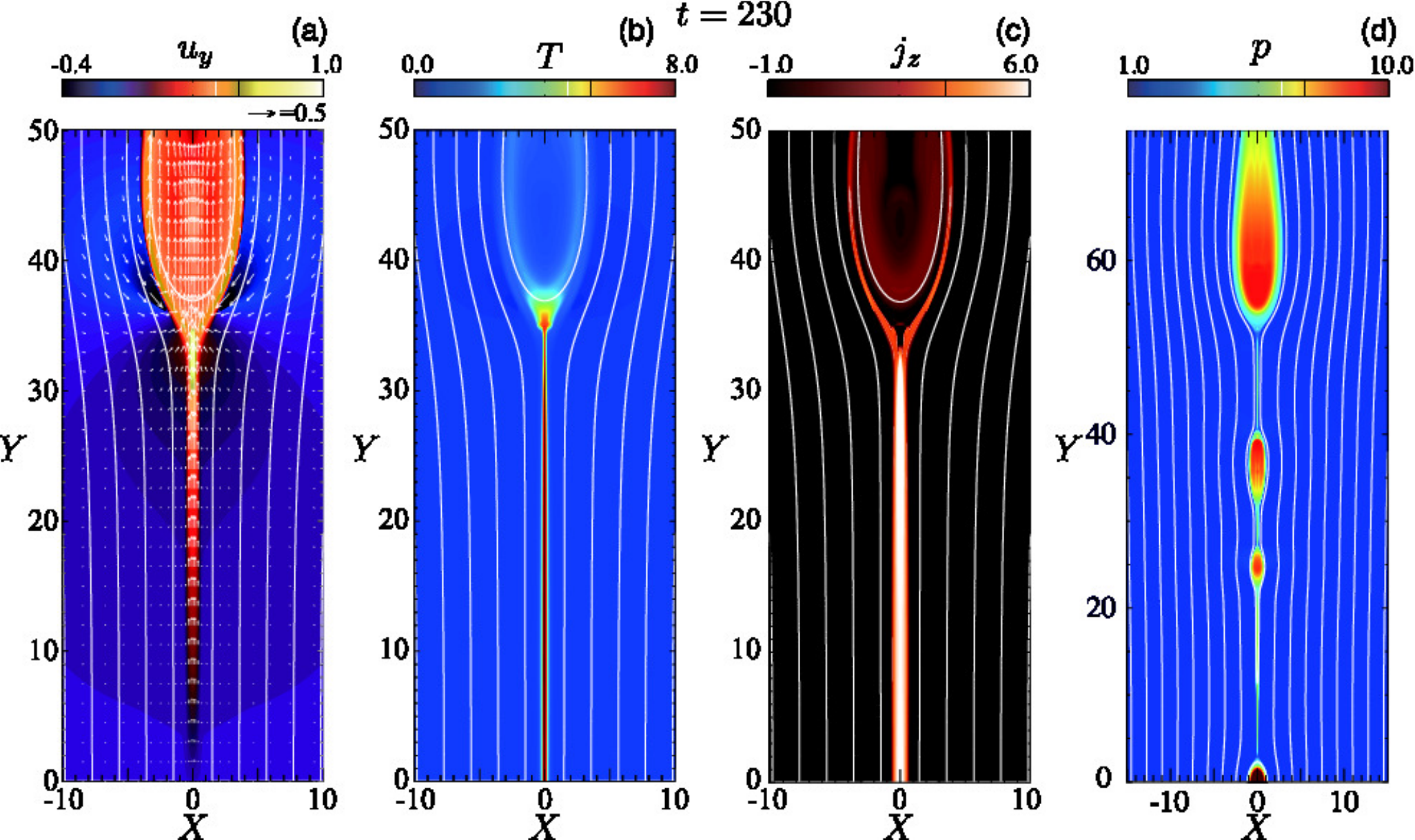}
\caption{
Sweet--Parker results from RRMHD simulations with a uniform resistivity in the reconnection plane. (a) The outflow component of the four-velocity ($U_y$), (b) the plasma temperature ($T = P/\rho$), and (c) the out-of-plane electric current density ($J_z$) are presented. The Lundquist number is $S \sim 10^{3.5}$. (d) The plasma pressure ($P$) with $S \sim 10^4$. The solid lines show the magnetic field lines. [Adapted from \citet{Takahashi2011}, reproduced by permission of the AAS]
\label{fig:takahashi11}
}
\end{figure}

\begin{figure}
\centering
\includegraphics[width=0.9\columnwidth]{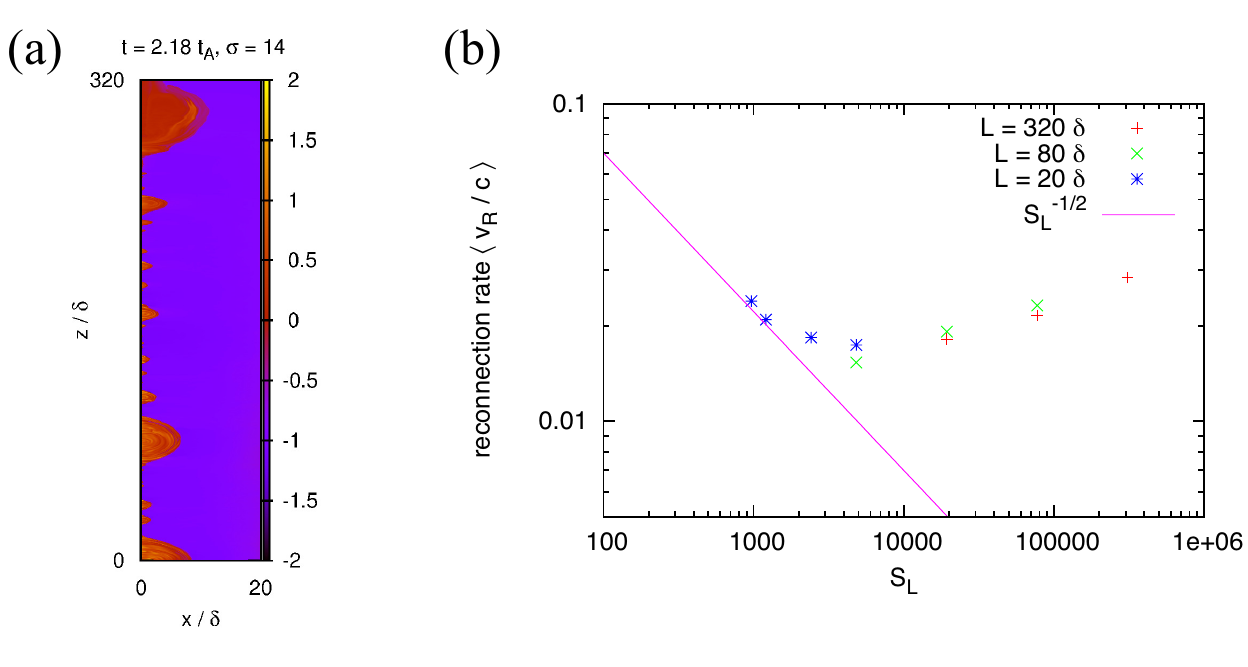}
\caption{
(a)
Plasma temperature $T/mc^2$ from an RRMHD simulation with $S \sim 10^{5.5}$ and  $\sigma=14$.
The figure shows half of the reconnection region, and the reconnecting magnetic field is along the z direction.
(b)
Time-averaged reconnection rate in the $\sigma=14$ runs,
as a function of the Lundquist number $S$.
The $L$ parameter indicates the length of the simulation box,
in unit of the initial current-sheet thickness $\delta$.
[Adapted from \citet{Takamoto2013}, reproduced by permission of the AAS].
\label{fig:takamoto13}
}
\end{figure}

\citet{Takahashi2011} have studied basic properties of Sweet--Parker reconnection.
They employed a uniform resistivity in Ohm's law (Eq. \ref{eq:ohm}), and then
they examined an RRMHD evolution of Sweet--Parker reconnection.
A laminar Sweet--Parker current sheet was successfully reproduced
in Figs. \ref{fig:takahashi11}(a)--(c).
By increasing the Lundquist number $S$ ($\propto \eta^{-1}$) from $2\times 10^3$ to $2\times 10^4$,
they have confirmed that the reconnection rate scales like $\propto S^{-1/2}$,
as predicted by the relativistic Sweet--Parker theory \citep{Lyubarsky2005}.
It was also reported that the outflow speed is sub-Alfv\'{e}nic,
because of the larger inertia by the relativistic enthalpy.

Similar to the nonrelativistic MHD reconnection,
when the Lundquist number exceeds $S \gtrsim \mathcal{O}(10^4)$,
the Sweet--Parker current sheet becomes turbulent
because of the repeated formation of plasmoids \citep{Biskamp1986,Loureiro2007,Bhattacharjee2009,Uzdensky2010}.
An early signature of the plasmoid-dominated regime
can be seen in Fig.~\ref{fig:takahashi11}(d), but
an RRMHD version of plasmoid-dominated turbulent reconnection has been studied by \citet{Takamoto2013}.
Fig.~\ref{fig:takamoto13}(a)
shows a representative result for $S\sim 10^{5.5}$.
One can see plasmoids of various sizes.
Fig.~\ref{fig:takamoto13}(b) presents
the $S$-dependence of the reconnection rate. 
The magenta line indicates the rate by the relativistic Sweet--Parker theory, which was numerically verified by \citet{Takahashi2011}.
For higher-$S$ regime of $S \gtrsim \mathcal{O}(10^4)$,
the system becomes plasmoid-dominated
the reconnection system becomes plasmoid-dominated.
As a result, the reconnection rate deviates from the Sweet--Parker rate, and it remains fast $\sim \mathcal{O}(0.01)$ regardless of $S$.

\subsection{Dependence to the resistivity model}\label{Dependence_to_the_resistivity_model}

\begin{figure}
\centering
\includegraphics[width=0.85\columnwidth]{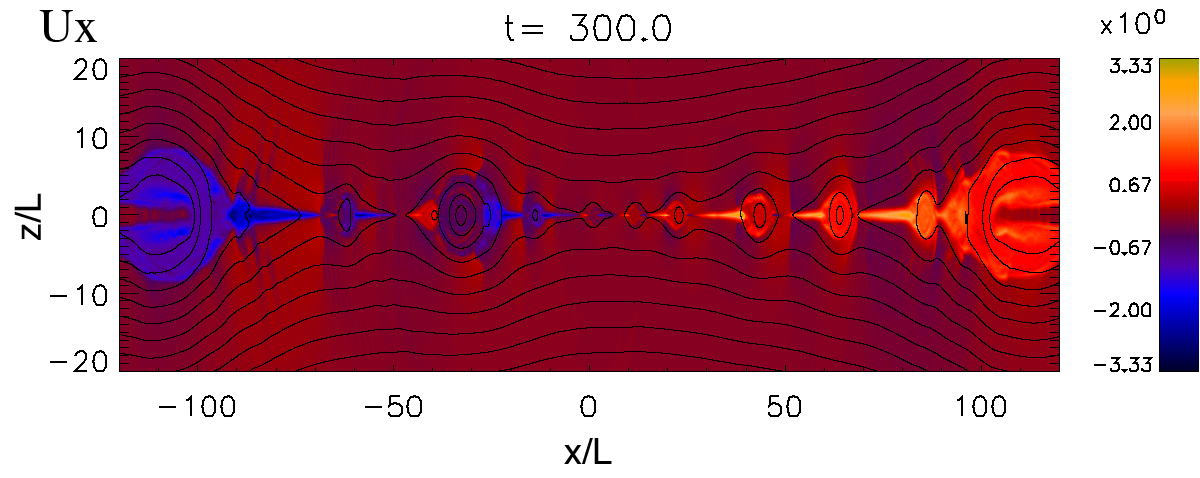}
\caption{
The $x$-component of the plasma 4-velocity $U_x=(\Gamma V)_x$ (outflow)
from a relativistic two-fluid simulation with uniform resistivity.
The black lines show magnetic field lines.
[Adapted from \citet{Zenitani2009a}, reproduced by permission of the AAS].
\label{fig:zeni09a}
}
\end{figure}

Similar to the nonrelativistic case,
the system evolution is sensitive to the effective resistivity model.
Relativistic Petschek reconnection develops
under a spatially-localized resistivity
in the RRMHD and relativistic two-fluid models \citep{naoyuki06,Zenitani2009a,Zenitani2010,Zanotti2011}.
In the RRMHD model,
Sweet-Parker \citep{Takahashi2011} and
plasmoid-dominated reconnections \citep{Takamoto2013}
develop
under a spatially uniform resistivity.
Interestingly,
when we employ a uniform friction parameter ($\tau$)
in the relativistic two-fluid model,
multiple plasmoids appear in the reconnecting current sheet,
as shown in Fig. \ref{fig:zeni09a}.
This differs from the RRMHD results
because the dissipation mechanism is no longer the same.
In the relativistic two-fluid model,
the fluid inertia terms also play a role similar to resistivity. 
Finally, under a current-dependent resistivity,
repeated formation of plasmoids is observed
in the RRMHD model \citep{Zenitani2010}. 
For practical applications of the RRMHD model,
an accurate form of parameter-dependent resistivity needs to be developed.

\subsection{Shocks in the reconnection system}

\begin{figure}
\centering
\includegraphics[width=\columnwidth]{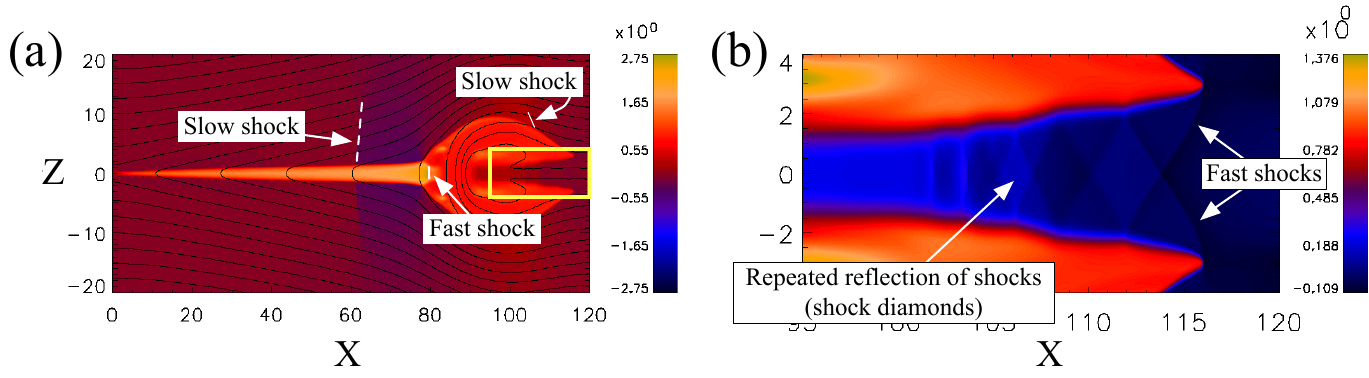}
\caption{
(a) A spatial profile of the $x$-component of the plasma 4-velocity $U_x=(\Gamma V)_x$ in the outflow direction from an RRMHD simulation with a localized resistivity.
The black lines show magnetic field lines.
(b) A profile of $U_x$ in the front side of the plasmoid, $x \in [95,120]$ and $z \in [-4,4]$ (corresponding to the yellow box in panel (a)).
[Modified from \citet{Zenitani2010},
reproduced by permission of the AAS].
\label{fig:zeni11b}
}
\end{figure}

We discuss another feature of relativistic magnetic reconnection in the fluid regime --- the system is often dominated by shocks. 
We remind the readers that
the relativistic sound speed is slower than $c/\sqrt{3}$.
In contrast, the outflow speed can be faster, approaching the speed of light $c$,
as the magnetization $\sigma$ increases in the upstream region (Eq.~\eqref{eq:Alfven}).
Comparing these relations, we find that
the outflow jet always becomes supersonic when $\sigma > 1/2$.
In such a supersonic regime,
the outflow jets and the outflow-driven plasmoids generate shocks.

In single-reconnection systems,
\citet{Zenitani2010} have reported various shocks
around the plasmoid ahead of the reconnection jet,
as indicated in Figs. \ref{fig:zeni11b}(a) and (b).
These shocks are essentially attributed to supersonic or transonic reconnection jets, but they need to be studied in further detail.
Even though \citet{Takamoto2013} has explored other important aspects,
no one has explored shocks in plasmoid-dominated systems.
We expect that
the plasmoid-dominated turbulent reconnection is also shock-dominated
in high-$\sigma$ regimes,
as potential signatures can be seen
in vertical discontinuities in Fig.~\ref{fig:zeni09a}.
To numerically deal with these shocks,
we need to use shock-capturing simulation codes. 
Most of the recent codes are capable of shocks
by using HLL-type upwind schemes \citep{Komissarov07,Palenzuela2009,Mizuno2013,Mignone2019}.

\subsection{Discussion}

These simulations in Sec.~\ref{sec2} have revealed
fluid-scale properties of relativistic magnetic reconnection.
Relativistic MHD reconnections are qualitatively similar
to nonrelativistic MHD reconnections --
relativistic Petschek \citep{naoyuki06,Zenitani2009a,Zenitani2010},
relativistic Sweet--Parker \citep{Takahashi2011}, and
relativistic plasmoid-mediated reconnections \citep{Takamoto2013} are reported.
The resistivity model (uniform vs localized) and inertial effects determine the system evolution,
as discussed in Section 2.5.
In these regimes, the system can be dominated by shocks.
At this point,
the number of RRMHD reconnection studies is still limited.
Many issues, such as the influence of environmental parameters, remain unsolved.
For example, dependence on
magnetization parameters: $\sigma \equiv B^2/w$ or $\sigma_m \equiv B^2/\rho$,
the effects of guide-field, flow-shear, and asymmetry need to be explored.

Beyond the special relativity, several groups have been actively developing
advanced Runge-Kutta codes for general relativistic resistive MHD (GR-RMHD)
\citep{Bucciantini2013,Dionysopoulou2013,Ripperda2019}.
\citet{Inda2019} have also developed an HLL-type GR-RMHD code to
study magnetic reconnection close to the black hole.
These GR-RMHD codes are successfully used to study the black hole and its accretion disk systems,
however, the influence of GR effects on the local reconnection physics remains unclear.
Combining radiative transfer equations with the RRMHD equations,
\citet{Takahashi2013} have developed
a relativistic radiative resistive MHD (RRRMHD) model.
The numerical results of RRRMHD reconnections look similar to the conventional RRMHD results,
however, the number of works is quite limited.
There is a strong demand for further studies,
to understand GR and/or radiation effects on RRMHD reconnection.

\section{Relativistic Reconnection in Collisionless Plasmas} 
\label{sec3}


While MHD simulations provide decent descriptions of the reconnection process over large scales, the diffusion region is likely collisionless in many active regions of interest. Two-fluid simulations describe part of the kinetic physics, but Particle-in-Cell (PIC) simulations can capture the full kinetics. Hence, in this section, we report the up-to-date progress in understanding relativistic reconnection using PIC simulations, which reveal the kinetic physics that breaks the ideal MHD condition and the key to fast magnetic reconnection in astrophysical collisionless plasmas.  

\subsection{Relativistic Generalized Ohm's Law}
\label{Relativistic_Ohms_Law}

During magnetic reconnection, magnetic flux is transported across the X-line. This requires the violation of the ideal condition
${\bf E} + ({\bf V}/c) \times {\bf B} = 0$, where ${\bf V}$ is the bulk plasma velocity. Understanding the physical mechanism that breaks the ideal MHD condition is one of the most important topics in reconnection physics, and the generalized Ohm's law is critical in determining such a mechanism, as also discussed in the non-relativistic limit {\color{blue} (Liu et al. 2024, this issue)}.

The extension of the generalized Ohm's law to the relativistic regime can be nontrivial, and different formalism was derived from the electron momentum equation \citep{hesse07a,Zenitani2018}. Here we discuss the latest form derived in \cite{Zenitani2018}. For simplicity, the speed of light is set to be $c=1$. We begin with the stress-energy tensor
\begin{equation}
T^{\alpha\beta} = \int f(u) u^{\alpha} u^{\beta} \frac{d^3u}{\gamma}.
\end{equation}
where $u^{\alpha} = (\gamma, \gamma {\bf v} )$ is the particle four-velocity and particle Lorentz factor $\gamma=1/\sqrt{1-(v/c)^2}$. A Greek index (e.g., $\alpha$, $\beta$) runs from 0 to 3 to account for four-dimensional spacetime. Using a four-velocity of the bulk flow $U^{\alpha}= (\Gamma, \Gamma {\bf V}$) where the fluid Lorentz factor $\Gamma=1/\sqrt{1-(V/c)^2}$, the metric tensor $g^{\alpha\beta}={\rm diag}(-1,1,1,1)$, and the projection operator $\Delta^{\alpha\beta} = g^{\alpha\beta} + U^{\alpha}U^{\beta}$,
the stress-energy tensor can be decomposed into \citep{Eckart1940}:
\begin{equation}
\label{eq:eckart}
T^{\alpha\beta}=\mathcal{E} U^{\alpha}U^{\beta} + q^{\alpha}U^{\beta} + q^{\beta}U^{\alpha} + P^{\alpha\beta}
\end{equation}
Here, $\mathcal{E} \equiv T^{\alpha\beta}U_{\alpha}U_{\beta}$
is the invariant energy density,
$q^{\alpha} \equiv -\Delta^{\alpha}_{\beta} T^{\beta\gamma} U_{\gamma}$
is the energy flux (heat flow), and
$P^{\alpha\beta} \equiv \Delta^{\alpha}_{\gamma}\Delta^{\beta}_{\delta}T^{\gamma\delta}$ is the pressure tensor.
The momentum part of Eq.~(\ref{eq:eckart}) is reduced to a familiar combination of the dynamic pressure and the pressure tensor,
${T}^{ij} = mn {V}^{i}{V}^{j} + {P}^{ij}$,
in the non-relativistic limit, where a Roman index (e.g., $i$, $j$) runs from 1 to 3 to account for the three-dimensional space.

Note that the choice of $U^\alpha$ in Eq.~(\ref{eq:eckart}) is arbitrary, and several choices can be considered (See \cite{Zenitani2018} for more detail).
Among them, here we employ the average plasma velocity ${\bf V}$ that carries the electric charge, and it thus satisfies the relation of $J_{\beta,(e)} = -e n' U_{\beta,(e)}$, as usual. Here $e$ is the unit charge, $n'$ is the proper number density, and $J_{\beta,(e)}$ is the electric current carried by electrons. We use the prime to denote the proper quantities in Sec.~\ref{sec3}.

We then use the energy-momentum equation of electrons,
\begin{equation}
\label{eq:T}
\partial_{\beta} T_{(e)}^{\alpha\beta}
= F^{\alpha\beta} J_{\beta,(e)}
= -e n' F^{\alpha\beta} U_{\beta,(e)}
\end{equation}
where $F^{\alpha\beta}$ the electromagnetic tensor. Subscript ${(e)}$ indicates electron fluid properties. However, for brevity, we omit this subscript hereafter in this subsection.

From the spatial parts (i.e., $\beta=1,2,3$) of Eqs.~(\ref{eq:eckart}) and (\ref{eq:T}), we obtain the electron Ohm's law,
\begin{eqnarray}
\label{eq:rel_ohm}
{\bf E}
=
- {\bf V} \times {\bf B}
- \frac{1}{\Gamma e n'}
\Big[
\partial_t T^{i0}
+
\partial_j( \mathcal{E} U^i U^j + Q^{ij} + P^{ij} )
\Big]
.
\end{eqnarray}
where ${\bf V}$ is the bulk velocity and ${Q}^{\alpha\beta} \equiv q^{\alpha}U^{\beta} + q^{\beta}U^{\alpha}$ is the heat-flow part of the stress-energy tensor. The relativistic effects appear in $\Gamma$, $\mathcal{E}$ ($> n'mc^2$), ${Q}$, and their time derivatives.

\begin{figure}[ht]
\centering
\includegraphics[width=9cm]{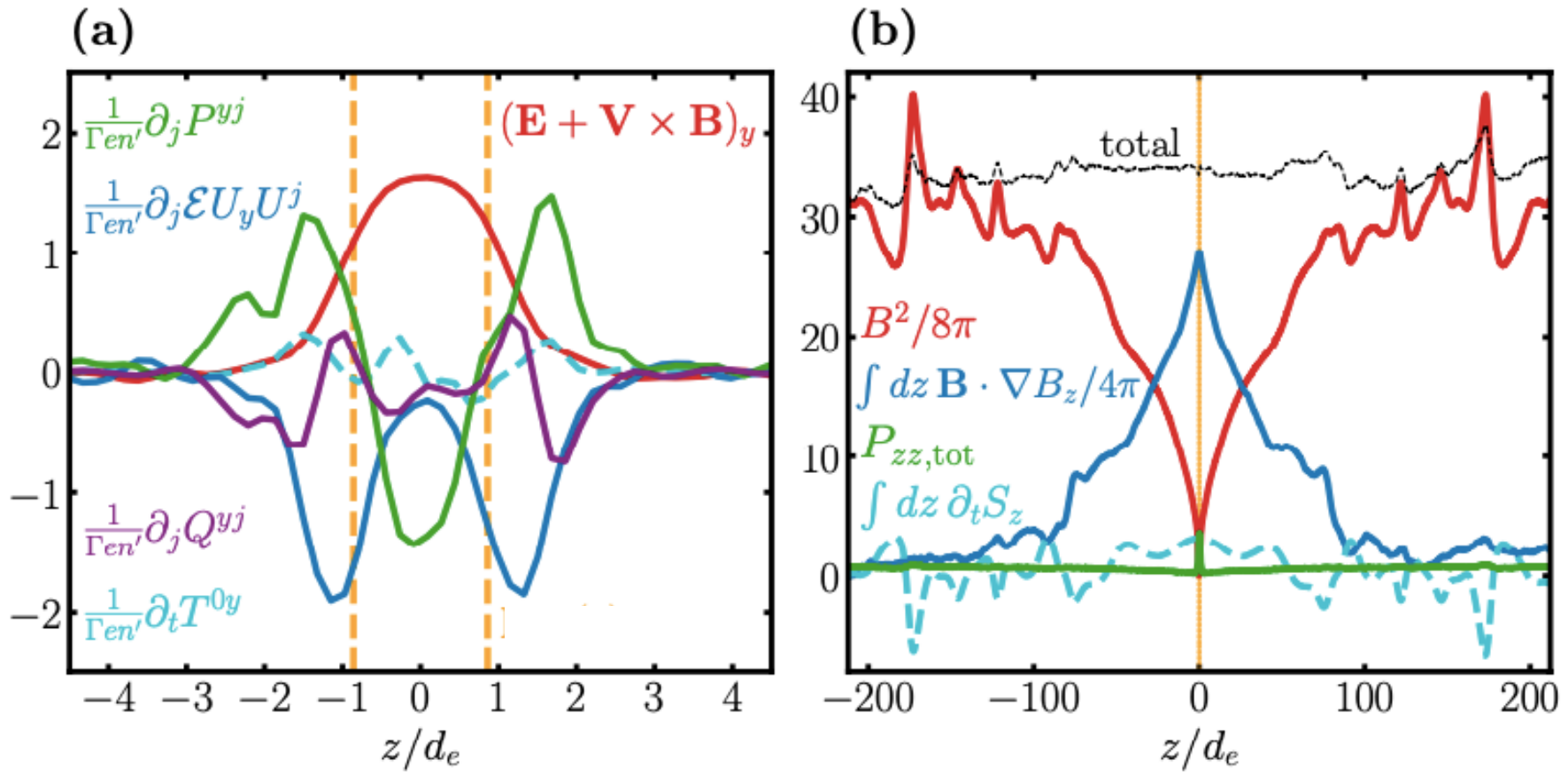} 
\caption {(a) The relativistic Ohm's law (Eq.~\ref{eq:rel_ohm}) in the ($y$-) direction out of the reconnection plane. Each term is color-coded. (b) The force balance along the inflow ($z$-) direction. The yellow vertical dashed lines in (a) and (b) mark the predicted diffusion region edges. [Adapted from \cite{Goodbred2022}]} 
\label{rel_Ohms}
\end{figure}

Figure~\ref{rel_Ohms} displays the out-of-plane ($y$-) component of the electron Ohm's law across the X-line in relativistic magnetic reconnection.
One key feature is the dominance of the inertial-like term $\partial_j( \mathcal{E} U_y U^j)$ in blue at the edge of the diffusion region where $({\bf E}+ {\bf V}\times {\bf B})_y$ in red is finite. This fact can be used to show that the diffusion region thickness is on the electron inertial scale \citep{Goodbred2022}, as marked by the yellow dashed vertical lines in Fig.~\ref{rel_Ohms}(a). This scale will be used to derive the first-principles reconnection rate in Section~\ref{localization_mechanism}.
The purple curve indicates the ``heat-flow inertial'' term,
which corresponds to the momentum transport by the energetic electrons that carry the heat flow in the rest frame of electrons \citep{Zenitani2018}. Relations between the kinetic physics and the terms in Eq.~(\ref{eq:rel_ohm}) are largely unknown and require more study, because these equations were formulated relatively recently.

The Eckart decomposition is also useful for evaluating the energy balance. The $0i$ components of the plasma stress-energy tensor in Eq.~(\ref{eq:eckart}) can be further decomposed into the matter flow, the bulk kinetic energy flux, the enthalpy flux, and the heat flux, as respectively shown in the right-hand side of the following equation,
\begin{eqnarray}
\label{eq:eflux}
T^{0i}
&=&
n'U^i + (\Gamma-1) n'U^i + \Big[ (\mathcal{E} - n')\Gamma U^{i} + P^{0i} \Big] + Q^{0i}.
\end{eqnarray}
\cite{Zenitani2018} reported that most incoming electromagnetic energy is converted into the relativistic enthalpy flux in the downstream region during relativistic magnetic reconnection, in agreement with RRMHD discussion in Section \ref{RPR}.

\subsection{Relativistic Collisionless Reconnection Rate}\
\label{rate_model}

We divide the reconnection rate problem in collisionless pair plasmas into two separate subsections, with one modeling the maximum plausible rate and another one modeling the key localization mechanism that leads to fast reconnection. Combining these two subsections, one can derive the reconnection rate as a function of magnetization from the first principles. Interested readers are encouraged to compare Sec.~\ref{rate_model} with the non-relativistic reconnection rate discussed in {\color{blue} Liu et al. (2024, this issue)}, which lays out the same approach to tackle the rate problem in kinetic current sheets of a wide variety of magnetic geometry, parameters and background conditions for heliophysics applications. 

\subsubsection{$R$-$S_{lope}$ relation and the maximum plausible rate}
\label{R_slope_relation}
As pointed out in earlier sections, in strongly magnetized plasmas, the plasma flow speed can be relativistic. During anti-parallel reconnection, the relevant force balance can be described by
\begin{equation}
\frac{({\bf B}\cdot\nabla){\bf B}}{4\pi}\simeq \frac{\nabla B^2}{8\pi}+\nabla\cdot{\bf P}+n'm_i({\bf U}\cdot \nabla){\bf U},
\label{force_balance}
\end{equation}
where ${\bf U}=\Gamma {\bf V}$, $\Gamma \equiv [1-(V/c)^2]^{-1/2}$ is the Lorentz factor of the bulk flow, and primed quantities are the proper quantities. Balancing the magnetic tension with the plasma inertia along the outflow direction, the resulting outflow speed is the relativistic Alfv\'en speed \citep{Blackman1994,Lyutikov2003,Lyubarsky2005,Zenitani2010,comisso14a,Liu2015} 
\begin{equation}
V_{A0}=c\sqrt{\frac{\sigma_{x0}}{1+\sigma_{x0}}}.
\label{VA0}
\end{equation}
which is limited by the speed of light when the cold magnetization parameter $\sigma_{x0}= B_{x0}^2/4\pi n' mc^2 \gg 1$. Here $B_{x0}$ is the asymptotic value of the reconnecting magnetic field component.

\begin{figure}[ht]
\centering
\includegraphics[width=11cm]{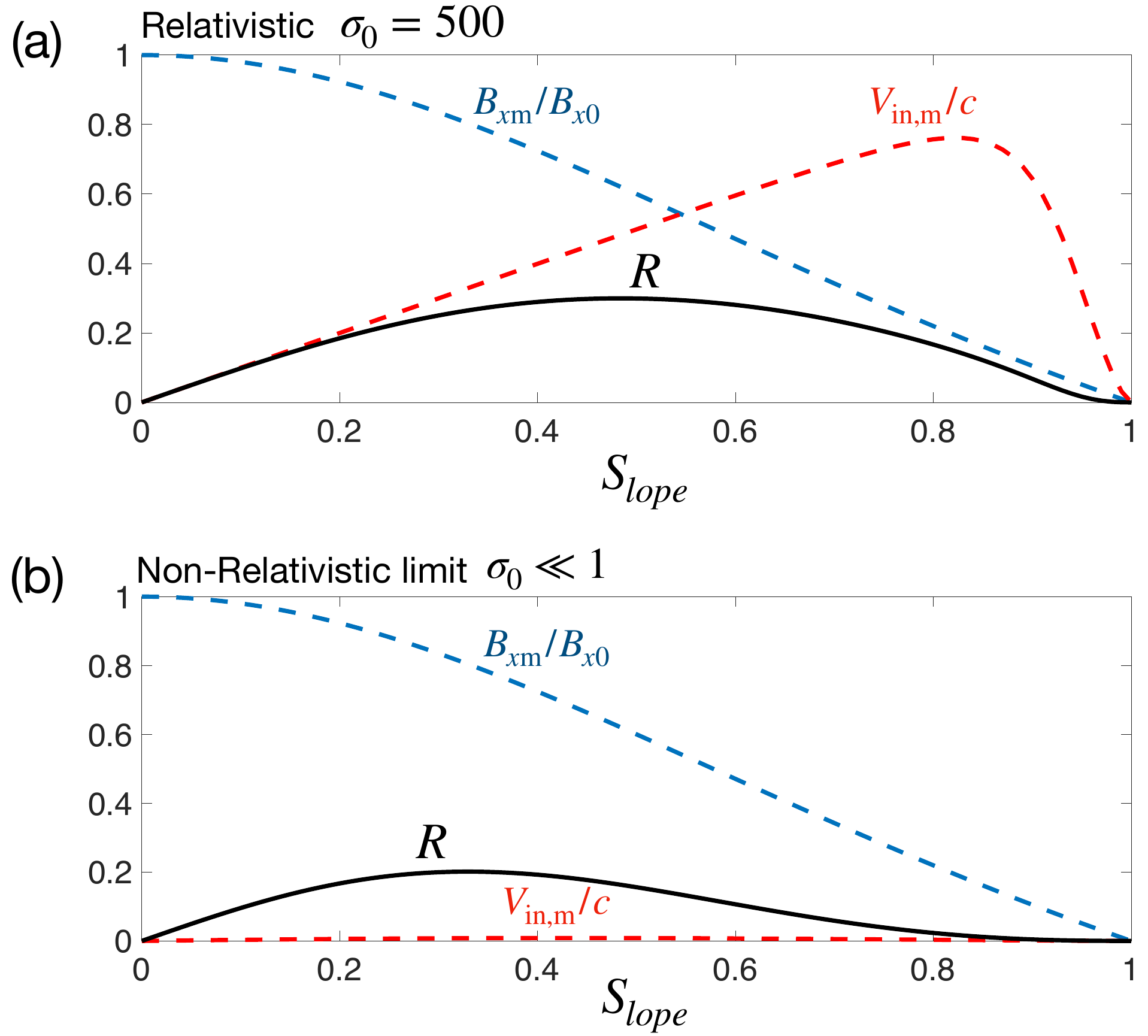} 
\caption {{\bf The $R-S_{lope}$ relation}. (a) The geometry and notation. (b) The predicted reconnection rate $R$, microscale inflow speed $V_{in,m}/c\simeq E_y/B_{xm}$, and field reduction $B_{xm}/B_{x0}$ as a function of separatrix $S_{lope}$ in the relativistic regime; (c) The predictions in the non-relativistic limit [Modified from \cite{Liu2017}]} 
\label{R_Slope}
\end{figure}

However, when the outflow geometry opens out, as shown in Fig.~\ref{R_Slope} (a), one needs to consider the difference in quantity magnitudes at the boundary of the microscopic diffusion region (denoted by subscript ``m'') and the mesoscale at the asymptotic region (denoted by subscript ``0''). Recognizing this scale separation is the key to obtaining an upper-bounded reconnection rate. By analyzing the force balance along the inflow (e.g., Fig.~\ref{rel_Ohms}(b)) and retaining the magnetic pressure ($\nabla B^2/8\pi$), \cite{Liu2017} derived
\begin{equation}\label{Bxm_Bx0}
\frac{B_{xm}}{B_{x0}}\approx \frac{1-S_{lope}^2}{1+S_{lope}^2}.
\end{equation}
where $S_{lope}$ is the slope of the separatrix that separates the reconnecting and reconnected field lines along the outflow exhaust boundary. Figure~\ref{R_Slope}(a) illustrates what these quantities mean. 
Similarly, analyzing the force balance along the outflow, retaining the magnetic pressure ($\nabla B^2/8\pi$), one can derive the outflow speed at the downstream boundary of the diffusion region 
\begin{equation}\label{Voutm}
V_{out,m}\simeq c \sqrt{\frac{(1-S_{lope}^2)\sigma_{xm}}{1+(1-S_{lope}^2)\sigma_{xm}}},
\end{equation}
which is smaller than $V_{A0}$ in Eq.~(\ref{VA0}).
Finally, the normalized reconnection rate  
\begin{equation}\label{rate}
R\equiv\frac{cE_y}{B_{x0}V_{A0}}=\left(\frac{B_{z{\rm m}}}{B_{x{\rm m}}}\right)\left(\frac{B_{x{\rm m}}}{B_{x0}}\right)\left(\frac{V_{out,{\rm m}}}{V_{A0}}\right),
\end{equation}
can be cast into a function of the separatrix $S_{lope}$ after plugging in Eqs.~(\ref{VA0}--\ref{Voutm}), and realizing that $B_{z {\rm m}}/B_{x {\rm m}}\simeq S_{lope}$ due to the geometry. 

As indicated by the blue curve in Fig.~\ref{R_Slope}(b), when the separatrix slope increases, the $B_{xm}/B_{x0}$ ratio decreases. Simulation \citep{Liu2017} suggests that in the high-$\sigma$ limit, the $B_{xm}/B_{x0}$ ratio can be significantly lower than that in the non-relativistic limit, which leads to a much higher microscopic inflow speed $V_{in,m}$ at the upstream boundary of the diffusion region, as predicted by the red curve in Fig.~\ref{R_Slope}(b). In spite of this relativistic inflow speed ($\sim 0.8c$), a value around $0.3$ still upper bounds the relativistic reconnection rate $R$. In comparison, the maximum plausible rate in the non-relativistic limit (Fig.~\ref{R_Slope}(c)) is around $0.2$, and the microscopic inflow speed is much lower than the speed of light. More discussion on the non-relativistic electron-proton plasmas can be found in {\color{blue} Liu et al. (2024, this issue)}.

\subsubsection{Localization mechanism that leads to fast reconnection}
\label{localization_mechanism}
\begin{figure}[ht]
\centering
\includegraphics[width=7cm]{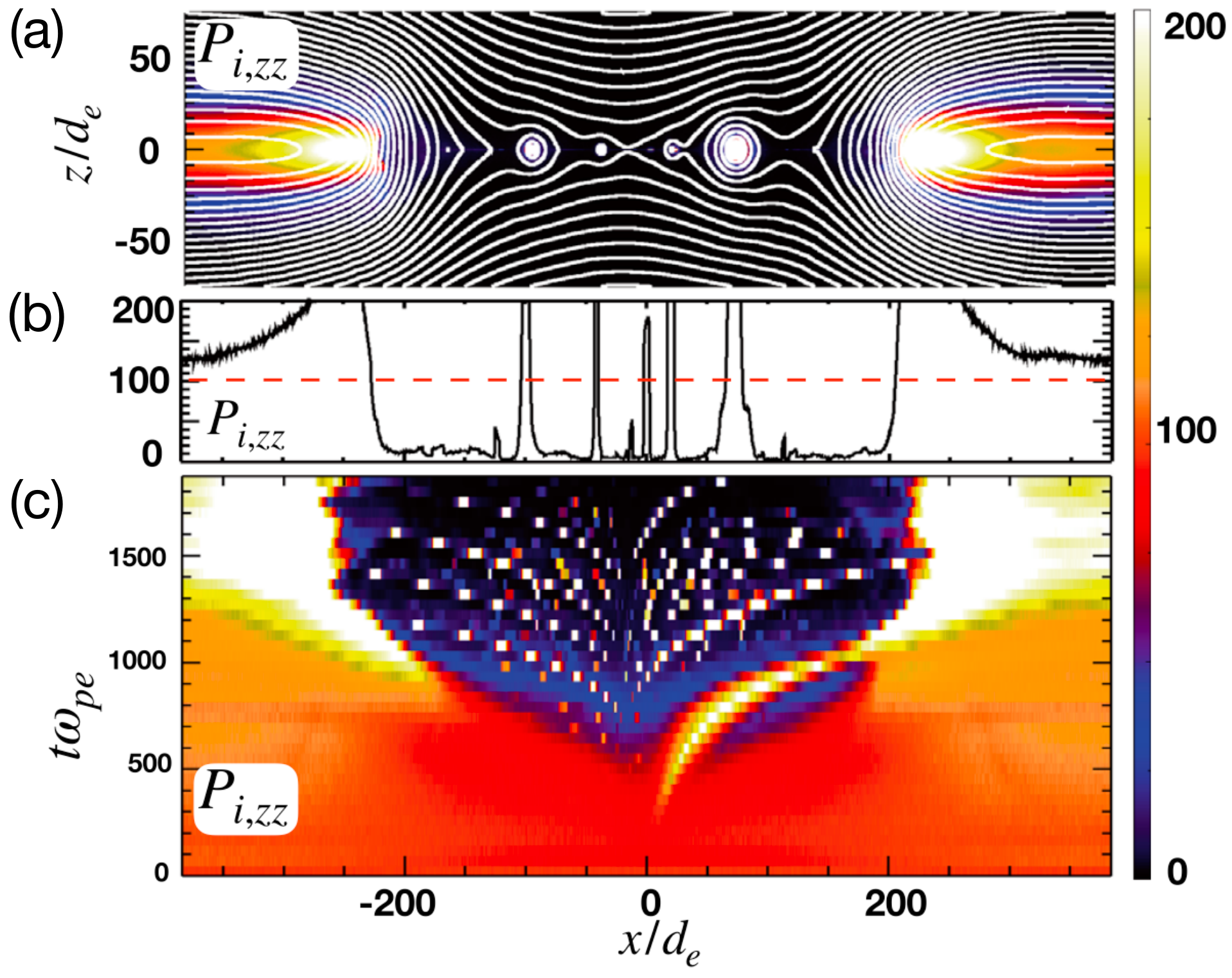} 
\caption {{\bf The pressure depletion and bursty nature of relativistic reconnection} with $\sigma_{x0}\simeq 89$. (a) The $P_{izz}$ contour. (b) The $P_{izz}$ cut along the outflow symmetry line (i.e., $z=0$). The red dashed horizontal line marks the initial plasma pressure. (c) The time stack plot of this cut.  [Adapted from \cite{Liu2020}, reproduced by permission of the AAS]} 
\label{bursty}
\end{figure}

In electron-positron (pair) plasmas, the Hall effect critical to facilitate fast reconnection in the electron-proton plasma \citep{sonnerup79a,mandt94a,shay99a,rogers01a,drake08a,yhliu14a,Liu2022} is absent, but the pressure depletion at the X-line appears to be significant, and it has an important consequence. This depletion is evident in Fig.~\ref{bursty}, which shows the positron pressure $P_{izz}$ contour in Fig.~\ref{bursty}(a) and its cut along the outflow symmetry line in Fig.~\ref{bursty}(b). The initial positron (also electron) pressure that can balance the asymptotic magnetic pressure is marked by the red dashed horizontal line of value $100 m_ec^2$, while the $P_{izz}$ cut at this nonlinear stage indicates a much lower value $\sim \mathcal{O}(m_e c^2)$ at the X-line (Note those peaks are secondary plasmoids that will be discussed later). This depleted pressure can cause the implosion of upstream plasmas into the X-line, providing the localization mechanism needed for fast reconnection in ultra-relativistic astrophysical plasmas. This idea is also illustrated by the cartoon in Fig.~\ref{R_Slope}(a).

\cite{Goodbred2022} managed to show analytically that the relativistic Lorentz factor associated with the large electric current density is responsible for this drastic pressure depletion. By considering the energy conservation along the narrow inflow channel toward the X-line, they derived the upper bound value of the X-line pressure,

\begin{equation}\label{eq:P2}
 P_{zz}\vert_{xline} \leq 2n'_{xline}mc^2 \left[\frac{\langle \gamma(v_z)\rangle_{xline}}{\Gamma_y}-\frac{\Gamma_y}{\langle \gamma(v_z)\rangle_{xline}} \right].
\end{equation}

\begin{figure}[ht]
\centering
\includegraphics[width=8cm]{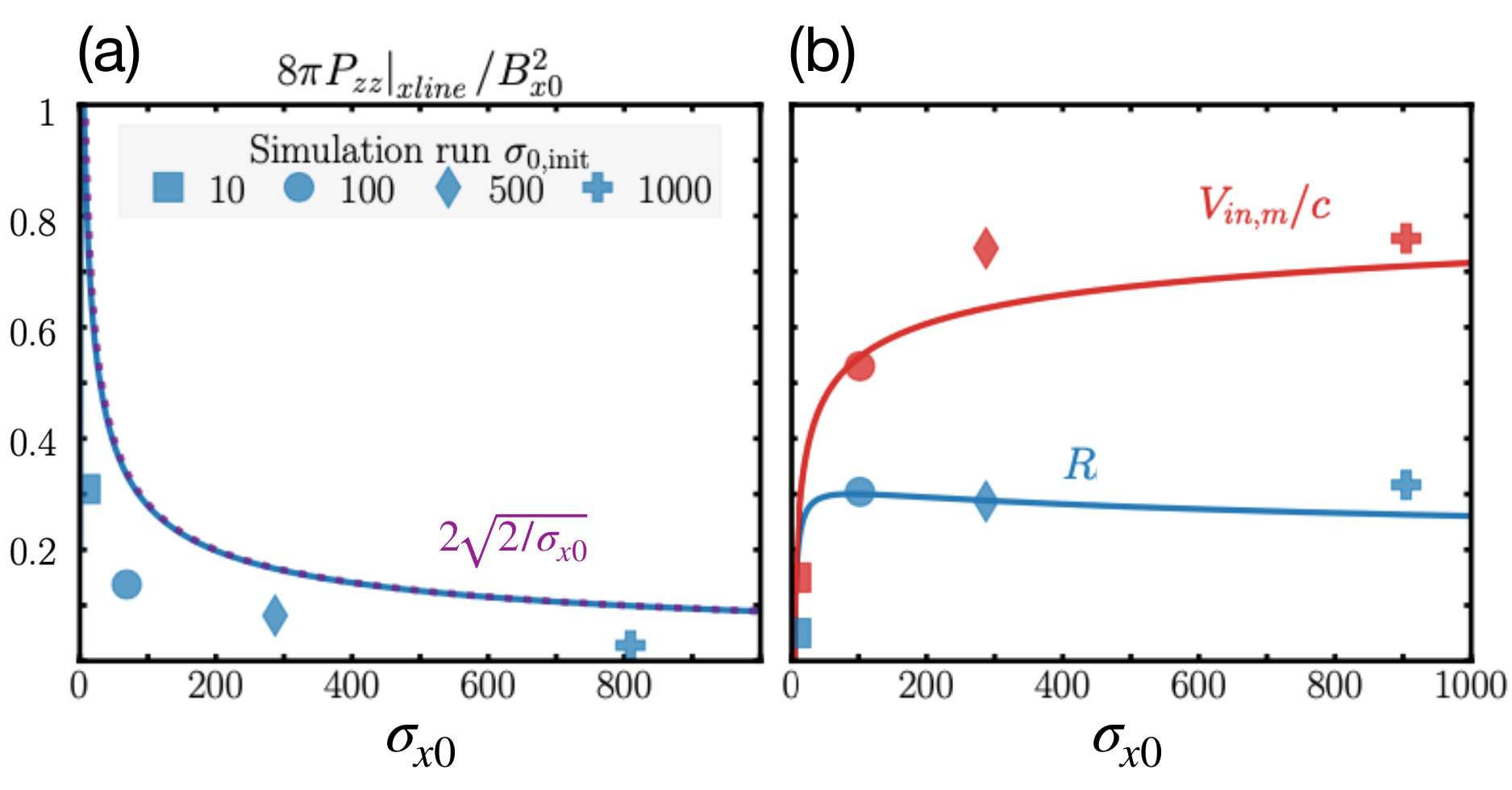} 
\caption {{\bf First-principles theory}. (a) The predicted upper-bound value of the X-line pressure (blue curve) as a function of $\sigma_{x0}$ that can be approximated as $2\sqrt{2/\sigma_{x0}}$ (purple dashed curve). The measured values in PIC simulations are shown as symbols. (b) The predicted reconnection rate (blue curve) and the microscale inflow speed $V_{in,m}$ (red curve) as a function of $\sigma_{x0}$. The measured values in PIC simulations are shown as symbols with the same color coding. Adapted from \cite{Goodbred2022}.} 
\label{rel_rate_theory}
\end{figure}

This expression clarifies the factors limiting the X-line thermal pressure. Here $\langle \gamma(v_z)\rangle_{xline}$ measures the $v_z$-averaged available energy at the X-line, and $\gamma$ is the Lorentz factor of a particle. On the other hand, $\Gamma_y\equiv [1-(V_y/c)^2]^{-1/2}$ only measures the bulk flow velocity of the current carriers that drift in the y-direction, whose magnitude is determined by the relativistic inertial scale that breaks the ideal MHD condition (Sec. \ref{Relativistic_Ohms_Law}). If all available energy is used to drive the current, then $\Gamma_y\simeq\langle \gamma(v_z)\rangle_{xline}$, and $P_{zz}\vert_{xline}$ becomes very small. Conversely, if only a small fraction of the total energy is needed to drive the current, then $\Gamma_y\ll \langle \gamma(v_z)\rangle_{xline}$ and $P_{zz}\vert_{xline}$ can become significant. The predicted X-line pressure normalized to the asymptotic magnetic pressure as a function of $\sigma_{x0}$ is shown in Fig.~\ref{rel_rate_theory}(a) as the solid lines, which scales as $\simeq 2(2/\sigma_{x0})^{1/2}$; It predicts a more severe pressure depletion in the large $\sigma_{x0}$ limit, capturing the decreasing trend of simulated X-line pressure shown by symbols.

Meanwhile, using the force balance along the inflow symmetry line within the diffusion region (ignoring the inflow inertia in Eq.~(\ref{force_balance})), one can relate the separatrix slope to the pressure difference between the X-line and upstream region, that is $P_{zz}\vert_{xline}$ (Eq.~(\ref{eq:P2})) in the cold upstream limit,
\begin{equation}\label{eq:slopes}
     S_{lope}^2 \approx 1-\frac{8\pi P_{zz}\vert_{xline}}{B_{xm}^2}.
\end{equation}

Since the upstream magnetic field line adjacent to the separatrix tends to straight out (when possible, due to the magnetic tension), we can couple this diffusion region solution with the larger upstream solution at the mesoscale (discussed in Sec.~\ref{R_slope_relation}) to get the reconnection rate. Namely, by plugging Eq.~(\ref{eq:P2}) into (\ref{eq:slopes}) to determine the $S_{lope}$,
then we can 
fully determine the reconnection rate from the $R$-$S_{lope}$ relation in Fig.~\ref{R_Slope}(b).
The prediction of $R$ is shown in Fig.~\ref{rel_rate_theory}(b) as the solid blue curve, which agrees well with the simulated reconnection rates shown as the blue symbols. 

\subsection{Bursty Nature of Relativistic Reconnection}
Most existing literature has focused on the application of plasmoid formation in astrophysics systems and its implication for particle acceleration. As to the origin of these plasmoids in PIC simulations, does it really resemble the ``high-Lindquist number plasmoid instability'' derived in the uniform resistivity MHD model \citep{Bhattacharjee2009,Loureiro2007,shibata01a}? The physics of the tearing instability can be different in collisionless plasmas. \cite{hoshino20a} shows that the collisionless tearing instability can actually be stabilized by the relativistic drift of current carriers, which was not captured in the resistive MHD model. Instead, in collisionless pair plasmas, the secondary plasmoids may be generated because the pressure within the reconnection exhausts is also depleted in the nonlinear stage, as shown in Fig.~\ref{bursty}(b); thus, plasmoids are violently produced from the collapse of the pressure-imbalanced current sheet. The formation of plasmoids within the primary exhausts helps balance the force, but only temporarily because they will be expelled out by outflows of the primary X-line. The evolution is settled into such a repetitive, dynamical balance, as shown in the time-stack plot in Fig.~\ref{bursty}(c).
These provide an alternative explanation to the bursty nature of relativistic reconnection in the antiparallel limit.  
It is interesting to note that a similar bursty nature was also reported in two-fluid simulation \citep{Zenitani2009a} in Fig.~\ref{fig:zeni09a} of Section~\ref{Dependence_to_the_resistivity_model}.
\cite{Liu2020} further demonstrated that the generation of secondary plasmoids can be suppressed in the presence of external guide fields, which is also not expected in the resistive-MHD model. As a potential application in plasma astrophysics, the mergers of secondary plasmoids are proposed to be a plausible generation mechanism of Fast Radiation Bursts (FRB) from neutron star magnetospheres \citep{Philippov2019,mahlmann22a}.

\subsection{3D Relativistic Turbulent Reconnection}

\begin{figure}[ht]
\centering
\includegraphics[width=\textwidth]{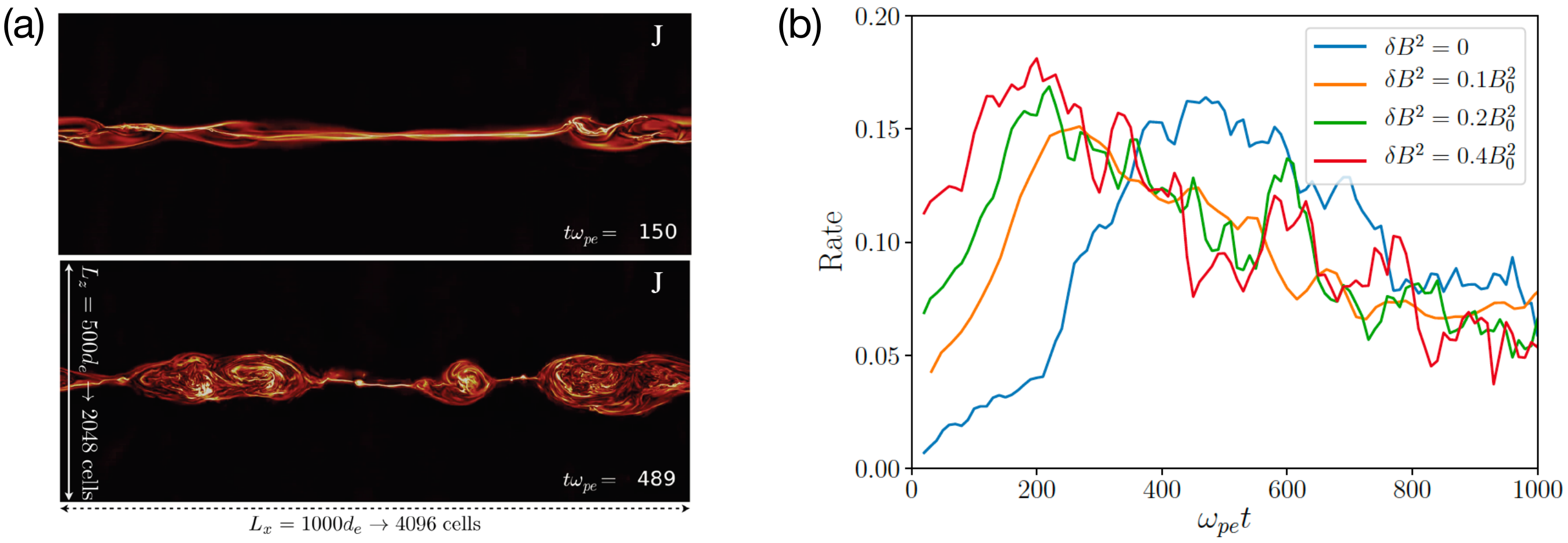} 
\caption {{\bf Relativistic turbulent 3D reconnection in electron-positron (pair) plasmas and its reconnection rates}. Panel (a) shows 2D cuts of current density during turbulent reconnection in the reconnection region at two different times; Panel (b) shows the evolution of the reconnection rate in cases with different initial turbulent fluctuation levels. Note that the rates are bounded by the predicted maximum plausible value ($\simeq 0.3$) shown in Fig.~\ref{R_Slope}(b) even with a strong fluctuation. [Adapted from \cite{Guo2021}, reproduced by permission of the AAS] } 
\label{turbulent_reconn}
\end{figure}

The discussion thus far in this section focuses on the features in 2D kinetic simulations. It is interesting to explore the differences and similarities in a 3D system where the current sheet coexists with background turbulence. Figure~\ref{turbulent_reconn} shows an example of relativistic turbulent reconnection in a 3D electron-positron plasma PIC simulation \citep[][also see Fig. \ref{fig-3D}]{Guo2021}. Fig.~\ref{turbulent_reconn}(a) shows the current density at two different times. The reconnection layer becomes highly structured because of the turbulent fluctuations initially imposed and later the self-driven turbulence arising from secondary oblique tearing instability \citep{Daughton2011} and flux-rope kink instability \citep{Zhang2021,Zhang2024}. Fig.~\ref{turbulent_reconn}(b) shows the evolution of the global reconnection rates in runs with different turbulence fluctuation levels. Interestingly, no matter how strong the imposed turbulence fluctuation is, the reconnection rate is still well-bounded by the value of 0.3, as predicted by the maximum plausible rate in Section~\ref{R_slope_relation}. Intriguingly, a similar result was demonstrated even in nonrelativistic resistive MHD simulations \citep{LYang20a} that constantly drive turbulence within the simulation domain. 
Since the turbulent reconnection in these simulations still develops large-scale, coherent inflows and outflows, we anticipate that the force balance and the global geometrical constraint still apply on average. Analytically, one can show that the prediction in Section~\ref{R_slope_relation} will work, as long as the outflow speed is on the order of ion Alfvén speed \citep{Liu2017}, no matter how thick \citep{SCLin20a} and complex \citep{yhliu18c} the diffusion region is. On the other hand, even though a thick turbulent diffusion region was theorized \citep{Lazarian1999}, the reconnection process in 3D PIC simulations, in fact, is still dominated by a few active diffusion regions in the kinetic scale, as also seen in Fig.~\ref{turbulent_reconn}(a). As long as the current sheet is in kinetic scale, the localization mechanism based on the fast drifting current carriers and pressure depletion, discussed in Section~\ref{localization_mechanism}, should also work; and this will lead to fast reconnection. Nevertheless, a full resolution to this complex setting remains largely unknown, and there are several other competing ideas on turbulent reconnection \citep{Lazarian1999,eyink11a,boozer12a,higashimori13a}; this topic continues to be an active research area of great interest.

\section{Plasma Heating and Particle Acceleration}\label{sec4}


Since magnetic reconnection in the magnetically dominated regime is likely associated with strong energy release, the heating and nonthermal particle acceleration during reconnection are of strong interest.
The past two decades have witnessed unprecedented progress in our understanding of nonthermal particle acceleration in relativistic magnetic reconnection. 
An important discovery is that relativistic magnetic reconnection supports strong particle acceleration and development of power-law energy distribution \citep{Zenitani2001,Sironi2014,Guo2014,Guo2015,Werner2016}. These discoveries are first found in the highly relativistic pair plasma and later in the mildly relativistic regime. This development also motivates the new studies in the nonrelativistic low-$\beta$ regime \citep{Li2019,Zhang2021,Zhang2024,Arnold2021}.

\subsection{Basic Acceleration Mechanisms}

The basic acceleration mechanisms during magnetic reconnection can be broadly categorized into two types. The ones that accelerate particles via non-ideal electric fields and the ones via the motional electric fields $\textbf{E}=-\textbf{V}\times\textbf{B}/c$. PIC simulations have uncovered several basic acceleration mechanisms such as Fermi-type acceleration, acceleration at X-line regions, and betatron acceleration, etc.  In addition, analytical theories have been proposed and built to understand the acceleration processes and the resulting energy spectra.

\begin{figure}[ht]%
\centering
\includegraphics[width=\textwidth]{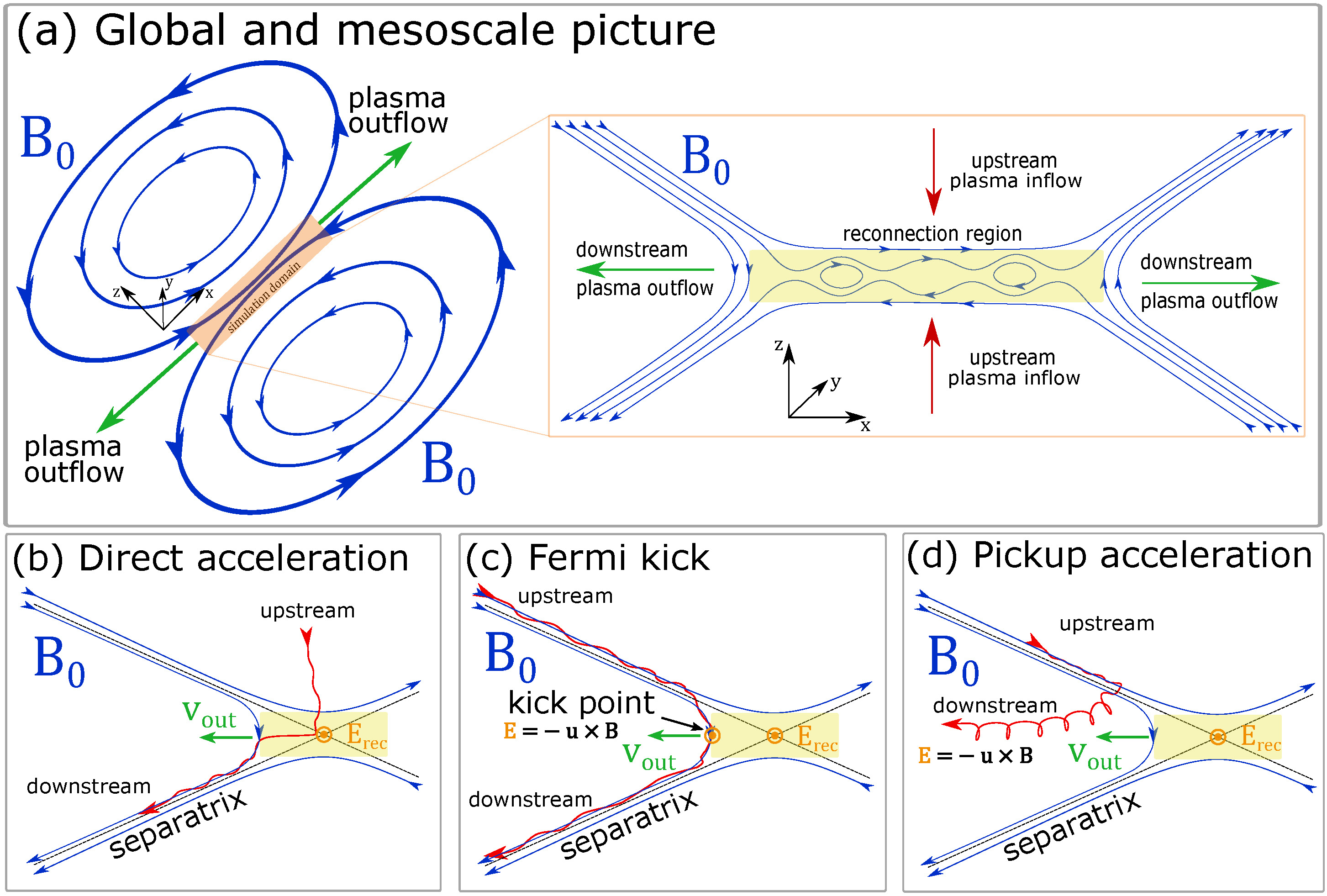}
\caption{Sketches of global and mesoscale reconnection configurations and several particle energization mechanisms [adapted from \citet{French2023}, reproduced by permission of the AAS]. (a) The surrounding astrophysical context of the reconnection region (highlighted in orange). (b) Direct acceleration from the reconnection electric field near an X-line. (c) Fermi acceleration in the exhaust region (d) Acceleration by the pickup process, where the particle becomes unmagnetized as it crosses the exhaust boundary. In panels (b-d), $B_0$ is the reconnecting magnetic field, $E_{\rm rec}$ is the reconnection electric field, and~$V_{\rm out} \simeq V_{Ax0}$ is the reconnection outflow speed, approximately equal to the in-plane Alfv\'en speed.}
\label{fig-basic-mechanism}
\end{figure}

Figure \ref{fig-basic-mechanism} shows a broad reconnection region where several acceleration mechanisms may occur. At X-lines, the ideal Ohm's law is broken, and a strong non-ideal electric field exists $E \sim R V_{A0} B_0$, where $R$ is the reconnection rate.
In relativistic magnetic reconnection without a guide field, X-line acceleration is often approximated by regions where the electric field is stronger than the reconnecting magnetic field $E>B$ \citep{Zenitani2001,Sironi2014}. However, significant acceleration has also been found in the broader region of the reconnection layer \citep{Zenitani2007,Guo2019}. \cite{Guo2014,Guo2015} proposed that Fermi acceleration due to curvature drift motions in contacting and merging magnetic lands (similar to \cite{Drake2006}, proposed in nonrelativistic reconnection) is important. They compared the acceleration of the parallel electric field and Fermi acceleration due to curvature drift acceleration, and found that Fermi acceleration plays a dominant role. While the whole reconnection domain may have a negative contribution of betatron acceleration due to the strong energy release (decaying magnetic field $\partial B/\partial t<0$), betatron acceleration may still be important in the converging islands \citep{Hakobyan2021}. In addition, the so-called pickup process, where particles become unmagnetized when entering the reconnection layer and gain energy in the outflow \citep{Drake2009,Sironi2020,French2023}, can be important for low-energy acceleration.

There have been recent efforts evaluating the relative importance of each mechanism \citep{Guo2014,Guo2015,Kilian2020,Sironi2022,Guo2023,French2023}. One possible way to distinguish different acceleration mechanisms is to decompose the electric field into the components perpendicular and parallel to the magnetic field ($E_\perp$ and $E_\parallel$) \citep{Dahlin2014,Ball2019,Kilian2020}, or the motional electric field and non-ideal electric field \citep{Guo2019}\footnote{As discussed in \citet{Lemoine2019}, any perpendicular electric field satisfying $\textbf{E}\cdot \textbf{B} = 0$ and $E^2 - B^2 < 0$ may support a generalized Fermi acceleration. The speed of the collision center is not necessarily the speed of the MHD flow.}. As mentioned above, the $E>B$ regions may better represent the X-line for a vanishing guide field. The analyses generally show that the acceleration associated with the motional electric field / perpendicular electric field (associated with Fermi/betatron or pickup process) dominates for a weak guide field and weakens for a stronger guide field \citep{French2023}, likely because of the compressibility of the layer \citep{Li2018}. The contribution of the perpendicular electric field also increases with the domain size. It is important to recognize that, during magnetic reconnection, most energy conversion is through a large-scale process by the magnetic tension release, rather than at small kinetic scales. Since the nonthermal particles take a large fraction of the released energy in relativistic reconnection, they must somehow `tap' the motional electric field. The mechanisms like Fermi acceleration would correspond to the motional electric field and, therefore, must be responsible for the main part of the nonthermal spectra.

Fermi acceleration can be more generalized in reconnection systems \citep{Hoshino2012b,Lemoine2019}. \citet{DalPino2005} and \citet{Drury2012} proposed a model considering the net compression in a reconnection layer and the effect of escape. Instead of Fermi acceleration by magnetic islands, here, particles are accelerated in the electric field induced by the reconnection inflow. Future studies are needed to show which Fermi acceleration is the most dominant one.

\subsection{Nonthermal Acceleration Uncovered by PIC simulations}

Early PIC simulations show that the energy spectrum near the X-line resembles a power-law distribution \citep{Zenitani2001}, whereas the energy distribution over the broader reconnection layer is much softer \citep{Zenitani2007}. 
Recent large-scale PIC simulations of relativistic magnetic reconnection show a power-law energy spectrum $f(\gamma-1) \propto \mathcal{E}^{-p}$ when integrated over the whole reconnection region with various spectral indices $p$ as hard as $p\sim 1$ \citep{Sironi2014,Guo2014,Guo2015,Melzani2014,Werner2016}. The nonthermal spectra have been also discussed for electron-proton plasmas, but with a reduced $\sigma$ as $\sigma \sim \sigma_i \sim \sigma_e/(m_i/m_e)$ \citep{Werner2018,Ball2018,Kilian2020,Li2023}. In this regime, the energy spectra became softer ($p \gtrsim 2$).
This trend naturally connects to the results from nonrelativistic reconnection, where the power-law spectra are much softer ($p \gtrsim 3-4$) \citep{Li2019,Arnold2021,Zhang2021,Zhang2024}. 

\begin{figure}[ht]%
\centering
\includegraphics[width=0.7\textwidth]{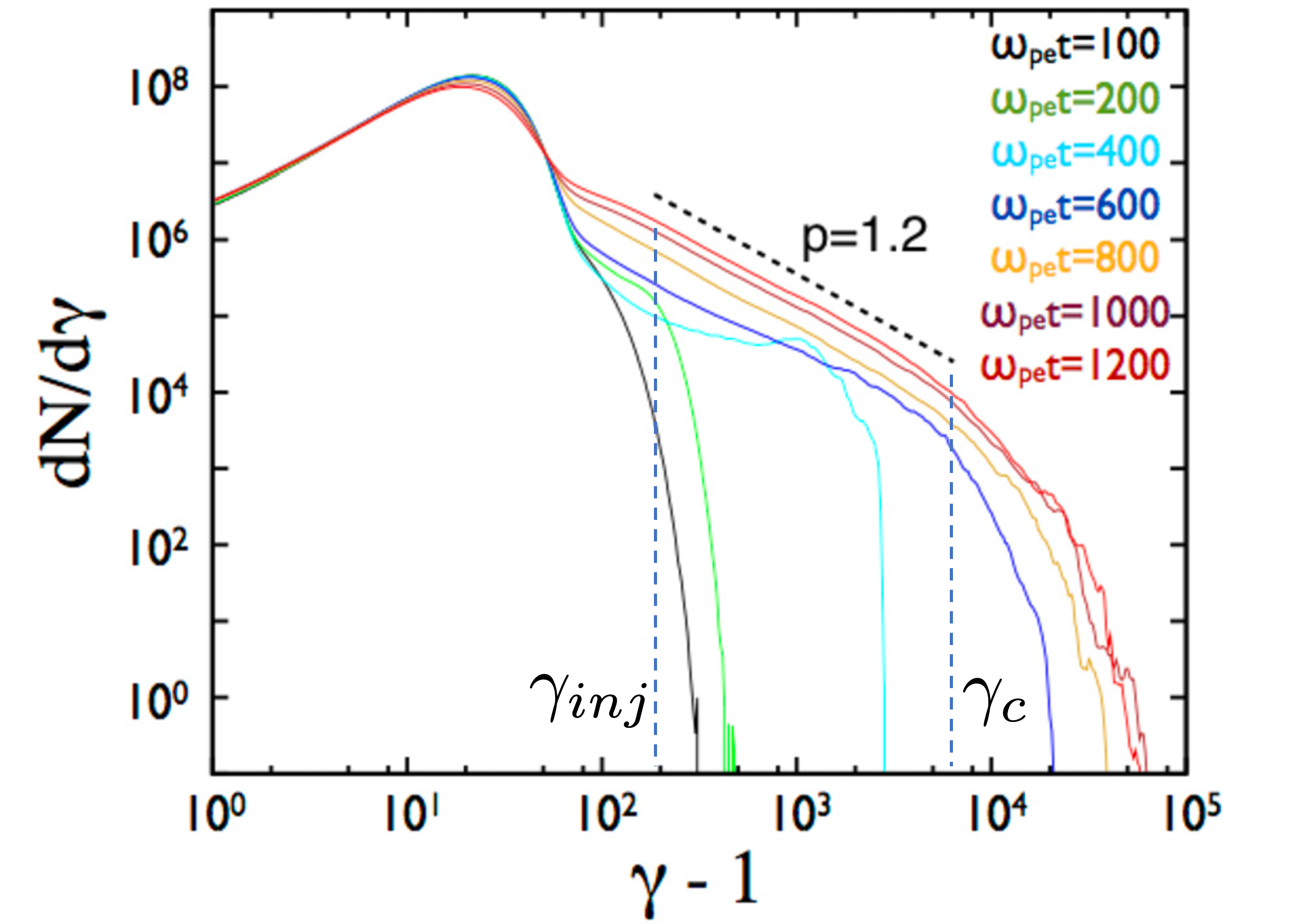}
\caption{The resulting energy spectrum in relativistic pair plasma reconnection
from a sample PIC simulation starting from a force-free current sheet with
$\sigma = 3200$ at different time steps generated
from the simulation (adapted and modified from \citet{Guo2020}.)}\label{fig-spectra}
\end{figure}

As shown in Fig. \ref{fig-spectra}, the nonthermal signatures found in PIC simulations can be understood in several aspects. A remarkably clear nonthermal power-law distribution is observed starting from a Lorentz factor a fraction of $\sigma$ to high energy. We term this lower-energy bound as the injection energy $\gamma_{inj}$, above which particles are accelerated into the nonthermal energies. This transition energy is also important for understanding the partition between thermal and nonthermal distribution \citep{Hoshino2023,French2023}. 
As the guide field becomes stronger, the energy spectra are softer, and the high-energy cutoff is suppressed \citep{French2023}. Finally, the energy spectra roll over at high energy, and the cutoff energy $\gamma_{c}$ scales with the simulation domain and time \citep{Petropoulou2018,French2023}.

\subsection{Physics that determine the acceleration results}

The strong nonthermal features in the particle energy spectra indicate several key processes, including how particle energization transits from thermal to nonthermal energies (``injection''), how power-law energy spectra develop (``power-law formation''), and the high-energy extension of the power-law (``roll over''). Below, we review the progress in understanding how they determine the energy spectra.

\subsubsection{Formation of Nonthermal Power-law Energy Spectra} 

Can magnetic reconnection support a clear nonthermal power-law spectrum, and if so, by what mechanism and under what condition? This has been a major theoretical issue in the particle energization during magnetic reconnection \citep[e.g.,][]{Drake2013,Guo2014}.  

\citet{Zenitani2001} proposed a simple power-law model.
Around the reconnection site, a particle is directly accelerated by the reconnection electric field $E_{\rm rec}$
in the out-of-plane direction, perpendicular to the reconnection plane.
Then its energy gain is approximated by
\begin{align}
\frac{d \mathcal{E}}{dt} = eE_{\rm rec}c .
\label{eq:zeni01_direct_acceleration}
\end{align}
They also estimate the loss rate of particles.
Since particles travel through (relativistic) Speiser motion,
their typical time scale $\tau(\mathcal{E})$ in the reconnection site can be approximated by a quarter-gyration by a typical reconnected magnetic field $\bar{B}_z$.
Then the particle loss rate is estimated by
\begin{align}
\frac{1}{N}\frac{dN}{dt}
= -\frac{4}{2\pi}\left(\frac{e\bar{B}_z}{\gamma mc}\right)
= -\frac{2ce\bar{B}_z}{\pi\mathcal{E}}
\label{eq:zeni01_loss_rate}
\end{align}
A key point is that the loss rate is energy-dependent.
Because of larger inertia $\gamma m$, higher-energy particles are less likely to escape from the acceleration site.
Combining Eqs.~\eqref{eq:zeni01_direct_acceleration} and \eqref{eq:zeni01_loss_rate},
we see that the particle number density follows the power-law distribution,
\begin{align}
N \propto \mathcal{E}^{-({2\bar{B}_z})/({\pi E_{\rm rec}})}
\end{align}
Since $\bar{B}_z/B_0 \sim \mathcal{O}(0.1)$ and $E_{\rm rec}/B_0 \sim \mathcal{O}(0.1)$, we see that the power-law index is on order of $\mathcal{O}(1)$. Similar models have been further developed by \citet{Uzdensky2022} and \citet{Zhang2023}, including magnetic flux ropes/islands as an escape region.

Recently, several new works discussed the formation of power-law distributions in a broad reconnection layer. \citet{Sironi2014} have proposed that the power-law form is established as the particles accelerated at the X-lines with $E>B$. They argue that this process is essential for the power-law formation and determines the spectra index of the energy spectra. However, since Fermi/betatron acceleration is the dominant acceleration in the broad reconnection region, it is unclear how X-line acceleration can solely determine the formation of the power law. \citet{Guo2014,Guo2015} proposed that the power-law distributions are produced by a Fermi-like process and continuous injection. In general, one can evaluate a Fokker-Planck-like equation for a reconnection layer \citep{Guo2020,Li2021,Li2023}

\begin{equation}
  \partial_t f + \partial_\mathcal{E}(\alpha_\text{acc}\mathcal{E} f) =
  \partial_\mathcal{E}^2 (D_{\mathcal{E}\mathcal{E}} f) - \alpha_\text{esc}f
  + \frac{f_\text{inj}}{\tau_\text{inj}},
  \label{eq:F-K}
\end{equation}

\noindent where $\mathcal{E}=(\gamma - 1)mc^2$ is the kinetic energy, $\alpha_\text{acc}$ is the acceleration rate, $D_{\mathcal{E}\mathcal{E}} = D_0\mathcal{E}^2$ is the energy diffusion coefficient, $\alpha_\text{esc}\equiv\tau_\text{esc}^{-1}$ is the escape rate, $f_\text{inj}$ is the injected particle distribution, and $\tau_\text{inj}$ is particle injection time scale. $\alpha_\text{acc}=(\partial_t\mathcal{E} +\partial_\mathcal{E} D_{\mathcal{E}\mathcal{E}})\mathcal{E}^{-1} $ describes a combination of the first-order Fermi processes and the accompanying first-order term associated with second-order Fermi mechanisms. 

As reconnection proceeds, the ambient plasma continuously flows into the reconnection layer through an inflow speed $V_{in}$ and undergoes a selective injection process. For a simple case with $D_{\mathcal{E}\mathcal{E}} = 0$, and $\alpha_{acc}$ and $\alpha_{esc}$ independent of energy, the solution of Eq.~(\ref{eq:F-K}) naturally recovers the classical solution $p = 1 + 1/(\alpha_\text{acc}\tau_{esc})$. A power-law distribution can form from an upstream thermal distribution when $\alpha_\text{acc} \tau $ is large ($\tau$ is the duration of acceleration) \citep{Guo2014,Guo2015}. In the limit that $\alpha_\text{acc}$ is large, the spectral index approaches $p=1$, consistent with existing PIC simulations. It was usually argued that an escape mechanism is necessary for forming a power-law distribution. This statement is incorrect or at least misleading. The power-law distribution can still form even for the case with no escape term \citep{Guo2014,Guo2015}. The main physics for forming a power-law is due to the continuous injection and Fermi acceleration. However, it is still important to understand the escape term, as it determines the shape of the distribution. Other acceleration with different $\alpha_{acc}$ can, in principle, form a power-law as long as the combined spectral index $p$ does not depend on energy,
\begin{equation}
 p = 1 + \frac{1}{\alpha_\text{acc} \tau_{esc}} + \frac{\partial \ln \alpha_\text{acc}}{\partial \ln \mathcal{E}} = {\rm const}
\end{equation}

\citet{Hakobyan2021} has developed an analytical model based on betatron acceleration to explain the high-energy nonthermal part of the spectra. \citet{Li2023} measured the acceleration and escape rates in PIC simulations, and the numerical solution of Eq.~(\ref{eq:F-K}) achieved a nice agreement with the PIC simulation results.

\subsubsection{Energy Partition of Thermal and Nonthermal Particles in Reconnection}

Magnetic reconnection is known to be the most important mechanism not only for plasma thermalization up to the equivalent temperature of the Alfv\'{e}n velocity, but also for accelerating nonthermal particles whose energies exceed their thermal energies \citep[e.g.,][]{Birn2007,Zweibel2009,Hoshino2012,Uzdensky2016,Blandford2017}. It was known that non-negligible fraction of nonthermal particles is generated during reconnection through satellite observations of the Earth's magnetosphere and the solar atmosphere \citep[e.g.,][]{Oieroset2002,Lin2003} and PIC simulation studies for non-relativistic plasmas \citep[e.g.,][]{Hoshino2001,Drake2006,Pritchett2006,Oka2010}. Furthermore, as magnetic reconnection is a ubiquitous process in the plasma universe, reconnection is believed to occur in many high-energy astrophysical phenomena such as in pulsar magnetosphere, accretion disks, and magnetar \citep[e.g.,][]{Remillard2006,Done2007,Madejski2016,Kirk2004,Lyutikov2003}.  PIC simulation studies also revealed that relativistic reconnection whose Alfv\'{e}n velocity is close to the speed of light can effectively generate nonthermal particles with a harder power-law energy spectrum than that generated in non-relativistic plasmas \citep[e.g.][]{Zenitani2001,Zenitani2005a,Zenitani2005b,Jaroschek2004,Jaroschek2009,Cerutti2012a,Cerutti2012b,Cerutti2013,Liu2011,Sironi2011,Sironi2014,Guo2014}. Since then, regardless of whether the plasma is nonrelativistic or relativistic, magnetic reconnection has gained attention as a mechanism of nonthermal particle acceleration in various space and astrophysical sources.
However, the energy partitioning of thermal and nonthermal plasmas during magnetic reconnection is not understood.  The energy partition plays an important role in the dynamical evolution of magnetic reconnection.  

\begin{figure*}
\includegraphics[width=\textwidth]{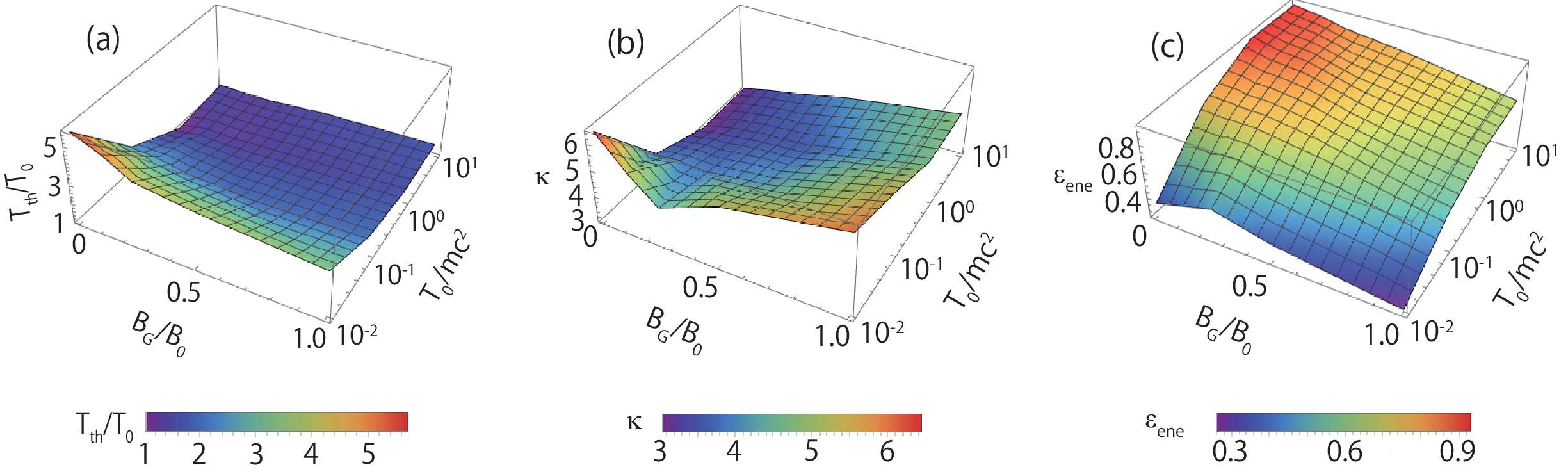}
\caption{Model fitting results as functions of the initial plasma temperatures $T/mc^2=10^{-2} \sim 10^1$ and the guide magnetic fields of $B_G/B_0=0 \sim 1$.  Three panels show (a) the average temperatures normalized by the initial background temperature $T_0$, (b) $\kappa$ index, and (c) the efficiency of nonthermal particles against thermal plasmas as a function of the initial plasma temperature $T/mc^2$ and guide magnetic field $B_G/B_0$. Adapted from \citet{Hoshino2023}, reproduced by permission of the AAS].}
\label{fig2}
\end{figure*}

Recently, the energy partitioning has been quantitatively investigated by using PIC simulations for a pair plasma \citep{Hoshino2022,Hoshino2023}.  The downstream heated plasma by reconnection has been found to be well modeled by a composed distribution function consisting of a Maxwell distribution function $N_M(\gamma)$ and a kappa distribution function $N_{\kappa}(\gamma)$, 
where
$N_M(\gamma) =n_M \gamma \sqrt{\gamma^2-1} \exp(-(\gamma-1)/(T_M/mc^2))$,
and
$N_{\kappa}(\gamma) =n_{\kappa} \gamma \sqrt{\gamma^2-1} \left(1 + (\gamma-1)/(\kappa T_{\kappa}/mc^2) \right)^{-(1+\kappa)} f_{cut}(\gamma)$, 
where $f_{cut}(\gamma)$ represents the high energy cutoff function given by $\exp(-(\gamma-\gamma_{cut})/\gamma_{cut})$ for $\gamma > \gamma_{cut}$.
\footnote{It is interesting to note that the kappa distribution function is widely observed in space plasma environments \citep{Vasyliunas1968,Livadiotis2013}, and the formation of the kappa distribution function may be related to the idea of non-extensive statistics \citep{Tsallis1988}.}
Based on the model fitting, Figure \ref{fig2} shows (a) the average thermal temperature of the Maxwellian and kappa distributions, which is described as $T_{th}=(n_M T_M + n_{\kappa} T_{\kappa})/(n_M + n_{\kappa})$, (b) $\kappa$ index, and (c) the fraction of the nonthermal energy density $\mathcal{E}_{\rm ene}$ as a function of the plasma temperature ($T_0/mc^2$) and guide field ($B_G/B_0$).  The nonthermal fraction $\mathcal{E}_{\rm ene}$ is defined as
$\mathcal{E}_{\rm ene}=
\int_1^{\infty} (\gamma -1) (N_{\kappa}(\gamma)-N_{\kappa}^{\rm M}(\gamma)) d \gamma /
 \int_1^{\infty} (\gamma -1) N_{{\rm M}+\kappa}(\gamma) d \gamma,$
 where $N_{\kappa}^{\rm M}(\gamma)$ represents the portion of the Maxwellian distribution function in the $\kappa$ distribution function.
For anti-parallel magnetic field topology without a guide magnetic field, it was found that while the nonthermal energy density in relativistic reconnection can occupy more than $90 \%$ of the total kinetic plasma energy density, most dissipated magnetic field energy can be converted into thermal plasma heating in nonrelativistic reconnection. 
For magnetic reconnection with a guide field, it is found that the fraction of nonthermal particles $\mathcal{E}_{\rm ene}$ basically decreases with the increase of the guide field.  However, for non-relativistic reconnection with $T/mc^2 \ll 1$ and a weak guide field, the nonthermal fraction $\mathcal{E}_{\rm ene}$ is found to increase.

\begin{figure*}
\includegraphics[width=\textwidth]{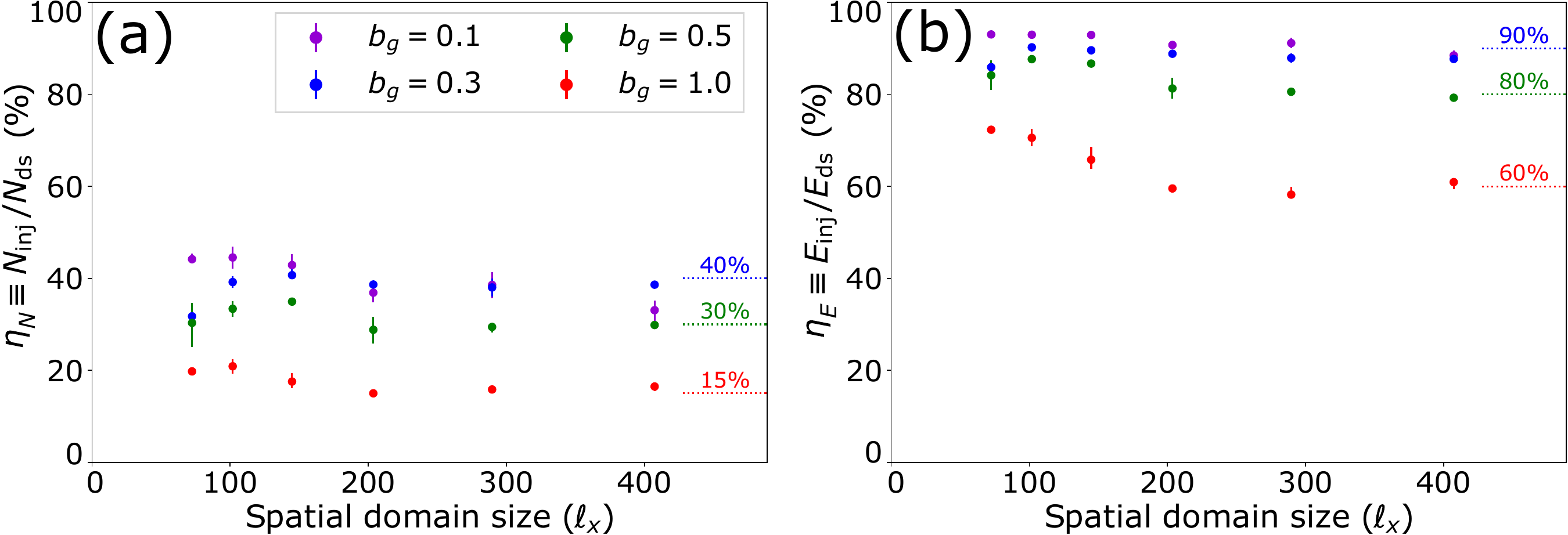}
\caption{Acceleration efficiencies from PIC simulations adapted from \citet{French2023}, reproduced by permission of the AAS. (a) The number efficiency (the ratio between the injected particle number to the downstream particle number) ~$\eta_N \equiv N_{\rm inj}/N_{\rm ds}$. (b) The energy efficiency~$\eta_E \equiv E_{\rm inj}/E_{\rm ds}$ for the ratio between the sum of injected particle energy to the energy of the whole downstream population).}
\label{fig-efficiency}
\end{figure*}

In the relativistic regime, a recent study by \citet{French2023} determines injection energy beyond which the particle energy distribution is well described by a power-law spectrum. They define the acceleration efficiency by the ``number'' and ``energy'', where the contribution of downstream nonthermal particles over the total downstream population is calculated. Consistently, they find that a stronger guide field suppresses the acceleration efficiency, and the efficiency may saturate to a final asymptotic value for a sufficiently large domain (see Fig. \ref{fig-efficiency}).

So far, these studies of energy partitioning have mainly been done in a two-dimensional system, but it is important to study a three-dimensional effect where turbulent magnetic reconnection can occur \citep{Daughton2011}.

\subsubsection{Injection problem}

\begin{figure*}
\includegraphics[width=\textwidth]{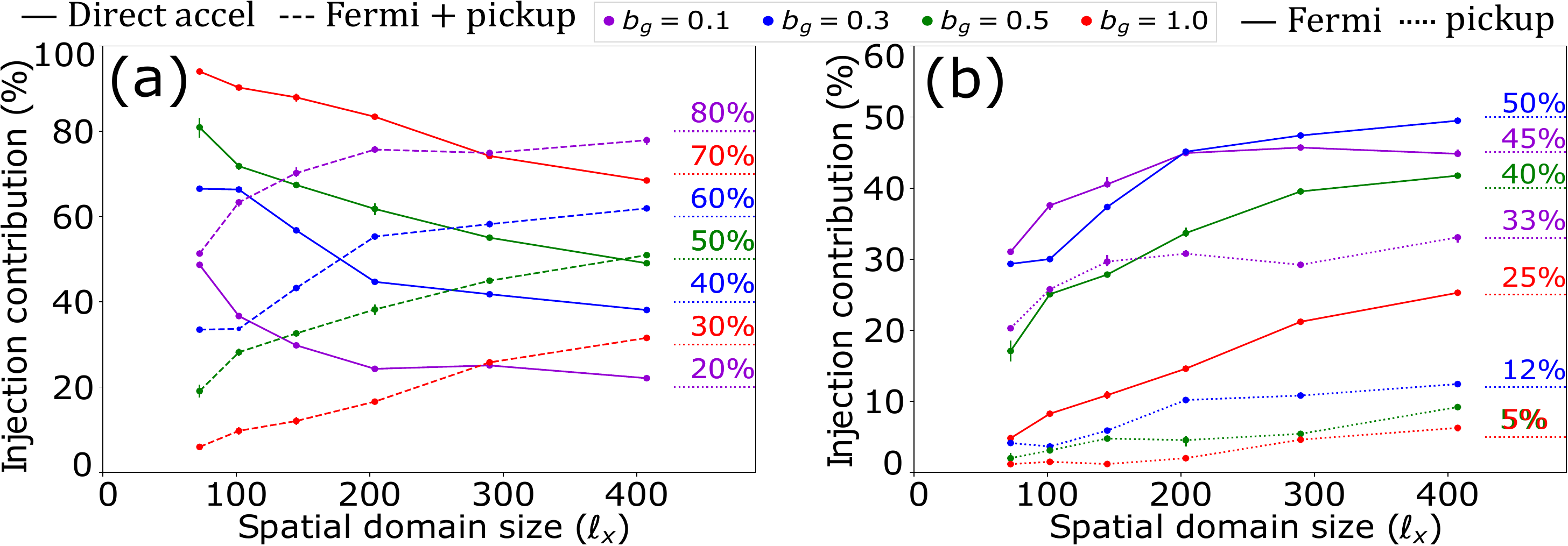}
\caption{Cumulative percentage of injected electrons that are injected by each particle acceleration mechanism (\citep{French2023}, reproduced by permission of the AAS]). (a): Decomposition between particles injected by~$W_\parallel$ (solid) and~$W_\perp$ (dashed). (b): Decomposition between Fermi-injected particles (solid) and pickup-injected particles (dotted). }
\label{fig-injection-mechanism}
\end{figure*}

There have been several studies on the injection problem with energization up until the injection energy $\gamma_{inj}$. While \citet{Ball2019} focused on the work done by the parallel electric field ($W_\parallel$), \citet{Kilian2020} studied the roles of both parallel and perpendicular electric fields. They showed that both parallel and perpendicular electric fields play a role, and perpendicular electric fields become more important for particle injection, especially when the simulation domain becomes large. While acknowledging the importance of $W_\perp$ during the development of the power-law distribution, \citet{Sironi2022} suggests that $E>B$ regions are important for injecting particles by accelerating particles, and they claimed all injected particles need to cross the X-lines. This apparent correlation is further studied by \citet{Guo2023}, and importantly, shown to have no significant contributions to the injection process. \citet{Guo2023} showed that $E>B$ regions contribute very little to injection ($\sim 10 \%\gamma_{inj}$) as they only host particles for a short time, insufficient for boosting the particles to injection. \citet{French2023} studied three different injection mechanisms (parallel electric field, Fermi reflection, and pickup process) and attempted to quantify their relative importance. Figure \ref{fig-injection-mechanism} shows that the Fermi and pickup processes, related to the electric field perpendicular to the magnetic field, govern the injection for weak guide fields and larger domains. Meanwhile, parallel electric fields are important for injection in the strong guide-field regime. \citet{Totorica2023} isolated the energy gain during injection from the nonideal field. They reached a different conclusion, because they focused on high-energy particles. To reach general consensus and to deepen our understanding, more research is needed on this important topic.

\subsubsection{High-energy roll-over}
Early results with a weak guide field show that the spectral index can be as hard as $p \sim 1$ for high $\sigma$, meaning the energy contained in such a spectrum is dominated by high-energy particles. The cutoff energy is therefore limited by the amount of dissipated magnetic energy $\gamma_c \sim [2\delta \sigma (2-p)]^{1/(2-p)}$, where $\delta$ is the fraction of the dissipated magnetic energy channeled into each species \citep{Guo2016}. For $p \sim 1$ this gives a high-energy roll-over at a few times $\sigma$ \citep{Werner2016}. It was suggested that as reconnection proceeds, the acceleration may evolve into softer spectra ($p\sim 2$), and the high-energy acceleration can proceed for a long time  \citep{Petropoulou2018,French2023}, although the $\sigma$ dependence still exists. Meanwhile, a stronger guide field or escape effects would make a difference.
A significant guide field leads to a softer spectrum and lowers the maximum energy \citep{French2023,Li2023}.
As reconnection proceeds, 
the escape process needs to be considered to correctly understand high-energy roll-over.

\subsection{The Roles of 3D reconnection and turbulence}

\begin{figure*}
\includegraphics[width=\textwidth]{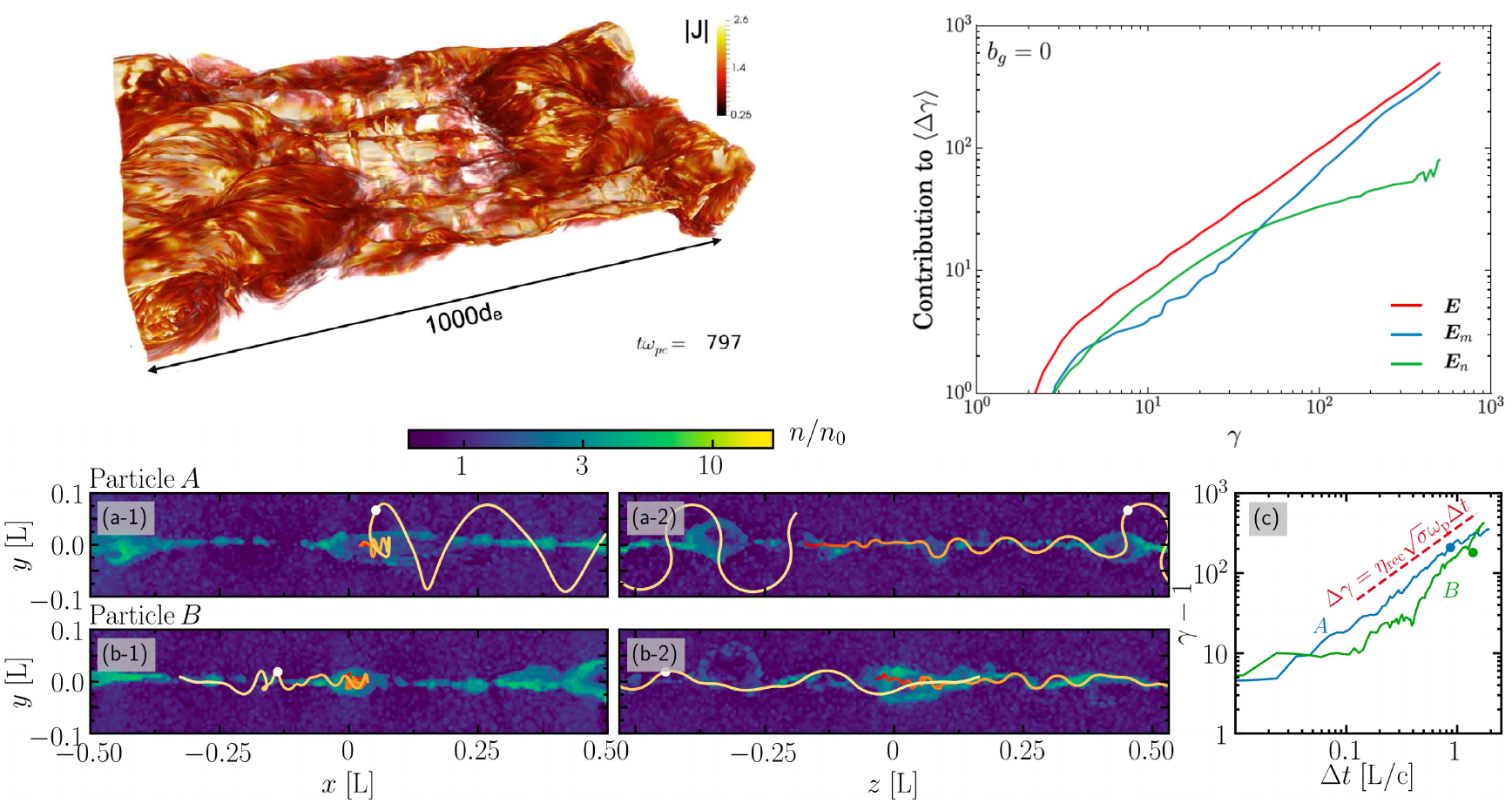}
\caption{Several recent studies on 3D relativistic magnetic reconnection and associated particle acceleration. Upper panels: The reconnection layer becomes significantly turbulent as the flux rope kink instability grows. The acceleration to high energy is dominated by the motional electric field ($\textbf{E}_m$), and the non-ideal electric field ($\textbf{E}_n$) plays a subdominant role (adapted from \citep{Guo2021}. Lower panels: 3D relativistic reconnection supports additional acceleration pattern, as particles can meander across the reconnection and gain energy in the upstream electric field \citep{Zhang2021ApJ...922..261Z}. Figures reproduced by permission of the AAS.}
\label{fig-3D}
\end{figure*}

Earlier discussions on 3D relativistic reconnection primarily focused on the drift kink instability \citep{Zenitani2007,Liu2011}. However, the flux rope kink instability can also grow and drive turbulence \citep{Zhang2021,Guo2021}. Both of these can be suppressed by a strong guide field \citep{Zenitani2008}. In the strong guide field regime, the oblique tearing instability is likely dominant \citep{Guo2021}.  As instabilities evolve, it is well established that 3D reconnection can spontaneously generate magnetic turbulence, as shown in Figure \ref{fig-3D}. Recent 3D studies have shown that 3D effects can be important for efficient acceleration in reconnection \citep{Dahlin2017,Li2019,Zhang2021,Zhang2022}. In 2D magnetic field configuration, particles are trapped in magnetic islands due to restricted particle motion across magnetic field lines \citep{Jokipii1993,Jones1998,Giacalone1994,Johnson2022}, and high-energy particle acceleration can be prohibited. Chaotic field lines and turbulence due to the 3D evolution of the oblique tearing instability \citep[e.g.,][]{Daughton2011,Liu2013} and flux rope kink instability \citep[e.g.,][]{Guo2021,Zhang2021} make particles leave the flux rope and can lead to efficient transport of particles in the reconnection region, which is found to be important for further acceleration in the reconnection region \citep{Dahlin2017,Li2019}. Bottom panels of Figure \ref{fig-3D} show that particles can escape from flux ropes and get acceleration in the reconnection inflow regions via Speiser-like orbits \citep{Zhang2022}.

\subsection{The problem of scale separation and macroscopic approach}
\label{macro}
PIC simulations have developed to the point that it is very useful to study the elementary processes of relativistic magnetic reconnection and particle acceleration (see Shay et al. 2024 this issue), and they have been used to explore global physics in heliophysics applications. However, one should remember that the scale separation in astrophysical problems is far beyond the reach of PIC simulations. For example, the ratio between the system size and the plasma skin depth can be $\sim 10^{13} - 10^{17}$ for pulsar wind nebulae and extragalactic jets. The conclusions made by PIC simulations need to be extrapolated to a large scale, and it is unknown if all the conclusions can still hold on a large scale. The solution to this serious issue is to develop a large-scale model that contains basic acceleration physics learned from PIC simulations. 

At large scales where the domain of the reconnection layer becomes much larger than the gyromotion scale, one can often use the guiding-center approximation, which includes Fermi/betatron acceleration identified as primary acceleration mechanisms \citep{Guo2014,Dahlin2014,Li2017}. An alternative solution is to use energetic particle transport theory. \citet{Li2018} have shown that the acceleration in the reconnection layer can be described as compression and shear, which are the main acceleration physics included in the energetic particle transport theory \citep{Parker1957,Parker1965,Zank2014,leRoux2015}. This approach has been used to study large-scale reconnection acceleration \citep{Li2018b,Li2022}. In addition, Fermi acceleration can be studied by following the momentum changes of particles through a sequence of local frames where the local electric field vanishes \citep{Lemoine2019}. Monte Carlo simulations \citep[e.g.,][]{Seo2023} including this description may alleviate the complicity of the particle transport equations when the plasma flow becomes relativistic \citep{Webb1989}. 

PIC simulations show that nonthermal particles can take a significant energy of the system, and therefore it may be important to include the feedback of energetic particles. \citet{Drake2019} presented a set of equations where the guiding-center particles feedback in the MHD equations so the total energy of the system for the fluid and `hot' particles is conserved. The feedback is through the pressure tensor of the energetic particles in the fluid equations. \citet{Arnold2021} successfully showed magnetic islands in the reconnection layer accelerate particles via Fermi acceleration and lead to the development of power-law energy spectra.  This description has not included the effect of particle scattering in turbulence, which is expected to be important in 3D turbulent reconnection. \citet{Seo2024} developed a new computational method for including the backreaction of energetic particles as a pressure term, while the distribution of the energetic particles is evolved through Parker's transport equation \citep{Parker1965}. An alternative (equivalent) way to include the feedback is through the electric current carried by energetic particles \citep{Bai2015,Sun2023}.




\section{Final Remarks} \label{sec5}

This review summarizes recent progress in relativistic magnetic reconnection, focusing on its fluid description and simulations, collisionless reconnection physics, and particle energization. We remark on several directions that may become frontier problems and need further attention. 

One of the most important topics is 3D magnetic reconnection. Although many progress has been made in understanding 2D and 3D reconnection, the 3D evolution is still an open issue. At this point, it is unclear whether magnetic reconnection always proceeds through a thin layer on the kinetic or a broader resistive scale. It is widely believed that the coupling between magnetic reconnection/tearing instability and current-driven instabilities, such as drift-kink and flux-rope kink instabilities, play an important role in the formation of turbulent current sheets. When 3D turbulence broadens the kinetic layer, its subsequent evolution can be sensitive to the current sheet and ambient plasma parameters. The 3D system evolution inevitably influences the way particles are accelerated, and therefore we need to clarify whether particle acceleration in 3D is fundamentally different from that in 2D.
These aspects have not been settled, and future progress is still very much needed to gain further insights.

As will be discussed in another article ({\color{blue} Nakamura et al., 2024, this issue}), 
it is important to understand the onset mechanism of magnetic reconnection.
Kinetic simulations typically start from a thin and long Harris sheet
in relativistic reconnection problems, but
the current sheet is highly unstable to kinetic instabilities
which grow in a few gyro-periods.
Then, it is not clear whether the Harris sheet is the best initial condition.
In order to justify many earlier works, or in order to discover the best initial condition,
we demand more research on the formation and stability of the ``initial'' current sheet.

In the future, in order to study the magnetosphere or even larger systems, we need to simultaneously resolve
kinetic scales of magnetic reconnection near the central object
and the large-scale evolution of the entire system.
Numerical simulation across such many scales will be challenging,
in particular in astrophysical settings.
To deal with the cross-scale problem, several attempts have been made
to interconnect small-scale kinetic simulations and large-scale MHD simulations
\citep{Sugiyama2007,Usami2013,Daldorff2014} for nonrelativistic problems,
however, the number of these MHD-PIC models is very limited.
Unfortunately, no similar attempts have been reported in relativistic astrophysics,
partly because it is difficult to translate kinetic quantities into fluid quantities
as discussed in Section \ref{sec3}.
A lot more work is necessary to develop relativistic MHD-PIC models. As we have discussed in Section \ref{macro}, it is important to consider nonthermal particle acceleration in MHD models, studying and predicting energetic particles and their signatures in large-scale systems.

Beyond the description discussed in this review article, incorporating the radiation \citep{Jaroschek2009,Cerutti2013,Zhang2018,werner19a,Sironi2020,schoeffler23a} and other QED processes \citep{schoeffler19a}, such as pair production and annihilation, during reconnection is also important for one to delve into the rich physics in such extreme astrophysical plasmas. On the other hand, this endeavor could also help extract observable signatures for distant observers, constraining our understanding of relativistic magnetic reconnection. 

With recent breakthroughs, relativistic magnetic reconnection has become an important topic in reconnection studies and a key process for understanding astrophysical energy release, particle acceleration, and high-energy radiation. Although difficult, it is highly anticipated that transformative progress will be achieved in the near future in key areas (Section \ref{sec1.2}) of relativistic magnetic reconnection, reaching a more complete physics understanding, achieving a more advanced modeling capability, and better connecting with astrophysical observations.

\bmhead{Acknowledgments}
F. G. acknowledges the support from Los Alamos National Laboratory through the LDRD program, DOE Office of Science, and NASA programs through 80HQTR20T0073, 80HQTR21T0087 and 80HQTR21T0104, and the ATP program. The work by Y. L. is funded by the National Science Foundation grant PHY-1902867 through the NSF/DOE Partnership in Basic Plasma Science and Engineering and NASA 80NSSC21K2048.
The work by S. Z. was funded by Japan Society for the Promotion of Science (JSPS) KAKENHI, Grant No. 21K03627.
The work by M. H. was supported by JSPS KAKENHI, Grant Nos. 19H01949 and 20K20908.

\bmhead{Conflict of interest}
The authors have no conflict of interest to declare that is relevant to the content of this article.

\end{document}